\documentclass[%
 aip,
 amsmath,amssymb,
reprint,%
]{revtex4-1}

\usepackage{graphicx}% Include figure files
\usepackage{dcolumn}% Align table columns on decimal point
\usepackage{bm}% bold math
\usepackage{amsmath, amsthm, amssymb, mathrsfs}
\usepackage[utf8]{inputenc}
\usepackage[T1]{fontenc}
\usepackage{mathptmx}
\usepackage{etoolbox}

\def\Fmag{$F_{\mathrm{mag}}(Q)$}
\def\Gmag{$G_{\mathrm{mag}}(r)$}
\def\dmag{$d_{\mathrm{mag}}(r)$}

\makeatletter
\def\@email#1#2{%
 \endgroup
 \patchcmd{\titleblock@produce}
  {\frontmatter@RRAPformat}
  {\frontmatter@RRAPformat{\produce@RRAP{*#1\href{mailto:#2}{#2}}}\frontmatter@RRAPformat}
  {}{}
}%
\makeatother
\begin{document}

\preprint{AIP/123-QED}

\title[Magnetic PDF Data from Polarized Neutrons]{Magnetic Pair Distribution Function Data Using Polarized Neutrons and \textit{ad hoc} Corrections}
% Force line breaks with \\
\author{Benjamin A. Frandsen}
 \affiliation{Department of Physics and Astronomy, Brigham Young University, Provo, UT 84602, USA}%Lines break automatically or can be forced with \\
 \email{benfrandsen@byu.edu}
\author{Raju Baral}%
 \affiliation{Department of Physics and Astronomy, Brigham Young University, Provo, UT 84602, USA}%Lines break automatically or can be forced with \\
\author{Barry Winn}
 
\affiliation{%
Neutron Scattering Division, Oak Ridge National Laboratory, Oak Ridge, Tennessee 37831, USA
}%

\author{V. Ovidiu Garlea}
 
\affiliation{%
Neutron Scattering Division, Oak Ridge National Laboratory, Oak Ridge, Tennessee 37831, USA
}%

\date{\today}% It is always \today, today,

\begin{abstract}
We report the first example of magnetic pair distribution function (mPDF) data obtained through use of neutron polarization analysis. Using the antiferromagnetic semiconductor MnTe as a test case, we present high-quality mPDF data collected on the HYSPEC instrument at the Spallation Neutron Source using longitudinal polarization analysis to isolate the magnetic scattering cross section. Clean mPDF patterns are obtained for MnTe in both the magnetically ordered state and the correlated paramagnet state, where only short-range magnetic order is present. We also demonstrate significant improvement in the quality of high-resolution mPDF data through application of \textit{ad hoc} corrections that require only minimal human input, minimizing potential sources of error in the data processing procedure. We briefly discuss the current limitations and future outlook of mPDF analysis using polarized neutrons. Overall, this work provides a useful benchmark for mPDF analysis using polarized neutrons and provides an encouraging picture of the potential for routine collection of high-quality mPDF data.
\end{abstract}

\maketitle

\section{\label{sec:intro}Introduction}

Magnetic pair distribution function (mPDF) analysis of neutron total scattering data has recently emerged as a valuable tool for investigating local magnetic correlations in magnetic materials~\cite{frand;aca14,frand;aca15}. In analogy to the more familiar atomic pair distribution function (PDF) method~\cite{egami;b;utbp12}, the mPDF is obtained by Fourier transforming the total magnetic scattering, which refers to the scattering arising both from long-range magnetic correlations (resulting in magnetic Bragg peaks) and from short-range magnetic correlations (resulting in diffuse magnetic scattering). This yields the real-space, pairwise magnetic correlation function. The mPDF technique is most useful for the study of short-range magnetic correlations such as those in a correlated paramagnet or a quantum disordered magnet, for which the real-space mPDF can be easier to interpret and model than the corresponding diffuse scattering pattern in reciprocal space. On the other hand, conventional magnetic Bragg diffraction analysis will typically remain the preferred choice for the determination of long-range ordered magnetic crystalline states. Since the introduction of the mPDF technique in 2014, it has been applied to numerous systems with short-range magnetic correlations ranging from quantum magnets to functional magnetic materials~\cite{frand;prl16,frand;prb16,frand;prm17,kodam;jpsj17,roth;prb19,tripa;prb19,lefra;prb19,zhang;prb19,frand;prb20,frand;prb21,baral;matter22}.

For an isotropic powder sample of a typical magnetic material possessing localized spins that belong to a single magnetic species, the mPDF is given by~\cite{frand;aca14, kodam;jpsj17}
\begin{widetext}
\begin{align}
G_{\mathrm{mag}}(r)&=\frac{2}{\pi}\int_{Q_{\mathrm{min}}}^{\infty} Q\left(\frac{\left(\text{d}\sigma/\text{d}\Omega\right)_{\mathrm{mag}}}{\frac{2}{3}N_sS(S+1)(\gamma r_0)^2 [f(Q)]^2}-1\right)\sin{(Q r)} \text{d}Q \label{FT}
\\&=\frac{3}{2 S(S+1)}\left(\frac{1}{N_s}\sum\limits_{i \ne j}\left[ \frac{A_{ij}}{r}\delta (r-r_{ij})+B_{ij}\frac{r}{r_{ij}^3}\Theta (r_{ij}-r)\right] - 4\pi r \rho_0 \frac{2}{3} m^2\right) \label{fullfofr}.
\end{align}
\end{widetext}
The first equation defines the experimental mPDF, while the second equation shows how to calculate the mPDF for a given magnetic structure. Here, $Q$ is the magnitude of the scattering vector, $Q_{\mathrm{min}}$ is the minimum measured scattering vector (assumed to exclude the small-angle scattering regime), $\left(\mathrm{d}\sigma/\mathrm{d}\Omega\right)_{\mathrm{mag}}$ is the magnetic differential scattering cross section, $r$ is real-space distance, $r_0=\frac{\mu _0}{4\pi}\frac{e^2}{m_e}$ is the classical electron radius, $\gamma = 1.913$ is the neutron magnetic moment in units of nuclear magnetons, $S$ is the spin quantum number in units of~$\hbar$, $f(Q)$ is the magnetic form factor, $N_s$ is the number of spins in the system, $i$ and $j$ label individual spins~$ \mathbf{S_{\textit i}}$ and~$\mathbf{S_{\textit j}}$ separated by the distance~$r_{ij}$, $A_{ij}=\langle S^y_i S^y_j \rangle$, $B_{ij}=2\langle S^x_i S^x_j \rangle - \langle S^y_i S^y_j \rangle$, $\Theta$ is the Heaviside step function, $m$ is the average magnetic moment in Bohr magnetons (which is zero for anything with no net magnetization, such as antiferromagnets), and $\rho_0$ is the number of spins per unit volume. A local coordinate system is used for each spin pair to compute~$A_{ij}$ and~$B_{ij}$, given by $\hat{\boldsymbol{x}}=\frac{\boldsymbol{r_{\textit j}}-\boldsymbol{r_{\textit i}}}{\vert \boldsymbol{r_{\textit j}}-\boldsymbol{r_{\textit i}} \vert}$ and $\hat{\boldsymbol{y}}=\frac{\boldsymbol{S_{\textit i}}-\hat{\boldsymbol{x}}(\boldsymbol{S_{\textit i}}\cdot \hat{\boldsymbol{x}})}{\vert\boldsymbol{S_{\textit i}}-\hat{\boldsymbol{x}}(\boldsymbol{S_{\textit i}}\cdot \hat{\boldsymbol{x}})\vert}$. The generalization to magnetic materials with multiple types of magnetic atoms and/or nonzero orbital contributions to the magnetic moment is straightforward and given elsewhere~\cite{frand;aca14}. Generally speaking, a positive (negative) peak at a given distance corresponds to net parallel (antiparallel) orientation between spins separated by that distance, lending an intuitive interpretation to mPDF data. We note that in Eq.~\ref{FT}, the subtraction of unity from the term in parentheses amounts to the removal of the self-scattering contribution to the magnetic scattering cross section, which is desirable because the self-scattering contains no information about the correlations between distinct magnetic moments. Eq.~\ref{FT} is fully equivalent to the corresponding integral expression used in the definition of the atomic PDF.

One of the biggest experimental challenges for obtaining high-quality mPDF data is accurately measuring the magnetic scattering to a sufficiently large momentum transfer and separating it from the (typically much larger) nuclear scattering. For a typical experiment using unpolarized neutrons, this can be done either by fitting a structural model to the nuclear Bragg peaks and subtracting them out to leave just the magnetic scattering, or by subtracting a reference measurement taken at a temperature where no coherent magnetic scattering is present (e.g. at high temperature well above a magnetic transition) from a diffraction pattern collected at a lower temperature with nonzero magnetic scattering. If the atomic structure is identical at both temperatures, then the difference between the scattering patterns gives the magnetic scattering. However, both of these strategies can result in significant artifacts due to imperfect modeling of the nuclear Bragg peaks (or nuclear diffuse scattering if short-range structural correlations exist) or imperfect temperature subtraction, where even simple thermal expansion causes shifts in the nuclear Bragg peak positions that can be difficult to treat accurately. Another source of error in this strategy is the over-subtraction of magnetic scattering which could still be present in the form of incoherent paramagnetic scattering that follows a form factor decay as a function of $Q$.

An alternative approach is to make no attempt to separate the magnetic scattering from the nuclear scattering and instead simply generate the PDF from the combined diffraction signal following standard protocols. The atomic and magnetic structures can then be modeled together in real space~\cite{frand;aca15}. However, this only works if the total scattering data are suitable for conventional PDF analysis, meaning that this approach is typically suitable only for data collected on dedicated PDF diffractometers with access to large values of momentum transfer (typically more than 20~\AA$^{-1}$). Even then, the mPDF signal can often be one or more orders of magnitude smaller than the nuclear PDF signal, making it difficult detect above the noise. Furthermore, this approach treats magnetic scattering the same as nuclear scattering in the PDF data processing protocol, meaning that no attempt is made to normalize the magnetic scattering by the squared magnetic form factor, as should normally be done according to Eq.~\ref{FT}. This has the effect of twice convoluting \Gmag\ with the Fourier transform of the magnetic form factor, broadening the mPDF by approximately $\sqrt{2}$ times the real-space width of the electronic wavefunction giving rise to the magnetic moment. This ``non-deconvoluted mPDF'' (previously called the ``unnormalized mPDF''~\cite{frand;aca15}) takes the form~\cite{frand;aca15}
\begin{align}
d_{\mathrm{mag}}(r)&=\frac{2}{\pi}\int_{Q_{\mathrm{min}}}^{\infty} Q\left(\frac{\text{d}\sigma}{\text{d}\Omega}\right)_{\mathrm{mag}}\sin{(Q r)} \text{d}Q\label{eq;drexp}
\\&=C_1 \times G_{\mathrm{mag}}(r)\ast S(r) + C_2 \times \frac{\textrm{d}S}{\textrm{d}r},\label{eq;dr}
\end{align}
where $C_1=\frac{N_s}{2\pi}\left(\frac{\gamma r_0}{2}\right)^2 \frac{2}{3}\langle g\sqrt{J(J+1)}\rangle^2$ and $C_2=\frac{N_s}{2\pi}\left(\frac{\gamma r_0}{2}\right)^2 \frac{2}{3}\langle g^2J(J+1)\rangle$ are constants, $\ast$ represents the convolution operation, and $S(r)=\mathcal{F}\left\{f_{m}(Q)\right\}\ast \mathcal{F}\left\{ f_{m}(Q)\right\}$ (where $\mathcal{F}$ denotes the Fourier transform). The second term on the right of Eq.~\ref{eq;dr}, which arises from the self-scattering contribution to the magnetic differential scattering cross section, results in a peak at low $r$ (below approximately 1~\AA). The strategy of co-modeling the atomic and magnetic PDF data in real space requires working with this lower-resolution, non-deconvoluted mPDF signal. To keep the distinction clear, we will refer to \dmag\ as the non-deconvoluted mPDF and \Gmag\ as the proper mPDF, deconvoluted mPDF, or just the mPDF. 

Another challenge exists even in cases where the magnetic scattering has been cleanly separated from the nuclear scattering: accurately normalizing the magnetic scattering by the squared magnetic form factor prior to performing the Fourier transform. Because the squared magnetic form factor for most magnetic ions approaches zero between 5 and 8~\AA$^{-1}$, any noise or leftover nuclear scattering in this range will be greatly amplified during the normalization step, inevitably leading to high-frequency artifacts in the mPDF data. Often, these artifacts are sufficiently severe that it is preferable just to skip the normalization step altogether and settle for the non-deconvoluted mPDF~\cite{frand;aca15}; however, this comes with the cost of significantly reduced real-space resolution, so it is also not ideal.

Many useful mPDF studies have been carried out in spite of these experimental challenges. Nevertheless, the technique will have a greater impact if the data quality improves. Here, we report two developments in mPDF data collection and processing that help address the dual challenges of accurately isolating the magnetic scattering and reliably mitigating the artifacts introduced during the magnetic form factor normalization step. The first is using longitudinal polarization analysis with a polarized neutron beam to separate the magnetic, nuclear, and nuclear spin-incoherent cross sections directly~\cite{schar;pssa93}, with no need for temperature subtraction or modeling of nuclear Bragg peaks. The second is using a polynomial function to perform \textit{ad hoc} data corrections that minimize the real-space artifacts resulting from the form factor normalization~\cite{billi;jpcm13,juhas;jac13}. These are both well-established methods, but they are applied here to mPDF analysis for the first time. In the following, we briefly describe longitudinal polarization analysis and the procedure for performing the \textit{ad hoc} data corrections, and then provide examples of their application to mPDF data on the antiferromagnetic semiconductor manganese telluride (MnTe). Code for performing the data corrections is freely available as part of the \texttt{diffpy.mpdf} python package~\cite{frand;jac22} and in the Supplementary Information of this article. These developments represent significant progress toward routine collection of high-resolution mPDF data on a variety of magnetic materials of fundamental and technological interest.

\section{\label{sec:polarized}Longitudinal Polarization Analysis}
The method of \textit{XYZ} longitudinal polarization analysis of neutron scattering data collected with large solid-angle detector banks (also called 6-pt polarization analysis, or 10-pt polarization analysis in its more general form including out-of-plane scattering) allows unambiguous separation of nuclear, magnetic, and nuclear spin-incoherent scattering cross sections. This is accomplished by measuring the spin-flip and non-spin-flip cross sections for various directions of polarization of the incident neutron beam; detailed descriptions of the technique are found elsewhere~\cite{schar;pssa93,stewa;jac09,ehler;rsi13}. A small handful of neutron diffractometers and spectrometers have the capability to perform longitudinal polarization analysis, such as D7 at the Institut Laue Langevin~\cite{stewa;jac09}, DNS at FRM-II in Munich~\cite{hmlz;jlsrf15}, and HYSPEC at the Spallation Neutron Source (SNS) and Oak Ridge National Laboratory (ORNL)~\cite{zaliz;jpconfs17}. The former two instruments can access a maximum momentum transfer of $\sim$4.0 and 4.84~\AA$^{-1}$, respectively, while HYSPEC can reach just over 6~\AA$^{-1}$ for certain choices of incident energy and detector positions. Given the importance of maximizing the range of accessible momentum transfer for the Fourier transform, HYSPEC is currently the instrument with polarization analysis that is most suitable for mPDF analysis.

\section{\label{sec:corrections}\textit{Ad hoc} Data Corrections}
In any attempt to isolate the magnetic scattering intensity, normalize it by the squared magnetic form factor, and perform the Fourier transform to generate the experimental mPDF, errors will inevitably be introduced due to imperfect time-independent instrumental background corrections, interference from nuclear and nuclear spin incoherent scattering, inaccuracies in the measured or approximated magnetic form factor, statistical noise, etc. Because typical squared magnetic form factors approach zero between 5 and 8~\AA$^{-1}$, any errors present in this range of the data become particularly magnified if included in the Fourier transform, introducing nonphysical artifacts into the mPDF. These challenges are similar to those encountered when generating x-ray PDF data, where the x-ray scattering atomic form factor plays an analogous role as the magnetic form factor. The traditional approach in x-ray PDF analysis has been to apply corrections for each source of error manually~\cite{egami;b;utbp12}, but this can be an onerous task and is itself prone to errors~\cite{billi;jpcm13}, particularly for researchers new to the technique. As has been discussed in detail for x-ray PDF~\cite{billi;jpcm13,juhas;jac13}, \textit{ad hoc} corrections using a polynomial correction function can largely mitigate the errors with minimal human input. This \textit{ad hoc} approach is also highly automatable, allowing for rapid throughput PDF data processing using tools such as PDFgetX3~\cite{juhas;jac13}. Here, we have implemented the PDFgetX3 algorithm in a custom-built python program and applied it to magnetic scattering data to obtain high-resolution mPDF data with no manual corrections necessary. The python code has been made available as part of the \texttt{diffpy.mpdf} package~\cite{frand;jac22} and as a standalone program in the Supplementary Information. 

One technical note merits discussion here. As explained in Ref.~\onlinecite{juhas;jac13}, the polynomial correction function results in nonphysical contributions to the real-space PDF for $r$ values less than $r_{\mathrm{poly}} = \pi n / Q_{\mathrm{maxinst}}$, where $n$ is the degree of the polynomial and $Q_{\mathrm{maxinst}}$ is the maximum value of $Q$ to which the correction polynomial is fit. As long as $r_{\mathrm{poly}}$ is shorter than the nearest-neighbor distance, then the artifacts resulting from the \textit{ad hoc} corrections will be restricted to the $r$-region below the first peak in the PDF data. In practice, an appropriate value of $r_{\mathrm{poly}}$ is selected (typically below 1~\AA\ for x-ray PDF data), and a weighted average of the PDFs obtained using the nearest integer values of $n$ corresponding to the selected $r_{\mathrm{poly}}$ value is generated. For synchrotron x-ray PDF experiments with $Q_{\mathrm{maxinst}}$ values of 25~\AA$^{-1}$ or greater, typical values for $n$ range from 7 to 9. When applied to mPDF, $Q_{\mathrm{maxinst}}$ will often be much smaller, anywhere from 5 to 8~\AA$^{-1}$, so $n$ should likewise be smaller to keep $r_{\mathrm{poly}}$ acceptably short. For HYSPEC, where the largest accessible value for $Q_{\mathrm{maxinst}}$ is about 6~\AA$^{-1}$, typical values of $n$ are 3 and 4, resulting in $r_{\mathrm{poly}}$ values of 1.57~\AA\ and  2.09~\AA\, respectively. Although these values would be too large for typical atomic PDF data, they are acceptable for typical magnetic structures, where the nearest-neighbor distance between magnetic atoms is unlikely to be as short as 2.09~\AA.

\section{\label{sec:results}Application to Manganese Telluride}
We demonstrate the use of polarized neutrons and the \textit{ad hoc} data correction procedure to generate mPDF data for the antiferromagnetic semiconductor MnTe. This material has a hexagonal crystal structure (space group $P6_3/mmc$) with a magnetic ordering temperature of $T_{\mathrm{N}}=307$~K, below which the Mn$^{2+}$ spins order with parallel alignment within the \textit{ab} planes and antiparallel alignment between neighboring \textit{ab} planes, as has been established by previous neutron diffraction studies~\cite{kunit;jdp64,dsa;jmmm05,szusz;prb06}. Short-range antiferromagnetic correlations are known to survive well into the paramagnetic phase above $T_{\mathrm{N}}$, likely up to 900~K or higher~\cite{zheng;sadv19}. The short-range magnetism in MnTe was the subject of a recent mPDF study~\cite{baral;matter22} using data collected with unpolarized neutrons, together with three-dimensional 3D-$\Delta$mPDF~\cite{roth;iucrj18} data collected from a single crystal.

\subsection{\label{subsec:experiment}Experimental Details}
A large powder sample of MnTe ($\sim$8~g) was synthesized according to the procedure in Ref.~\onlinecite{baral;matter22}, pressed into a pellet of diameter 6~mm, and loaded into an aluminum sample can. A closed cycle refrigerator was used to control the temperature. The incident neutron energy used for the experiment was $E_i=28$~meV, and the Fermi chopper was operated at a frequency of 60~Hz. This configuration provided an energy resolution at the elastic position of 5.7~meV full width at half maximum (FWHM). The data were integrated over energy transfers between $-5$ and 5~meV. The incident beam at HYSPEC is polarized by a vertically focusing array of magnetically saturated Heusler (Cu$_2$MnAl) crystals. A Mezei flipper was used to reverse the polarization direction of the incident beam, and the polarization analysis of the scattered beam was accomplished with a 60$^{\circ}$-wide-angle multi-channel radial supermirror-polarizer array~\cite{zaliz;jpconfs17}. Spin-flip (SF) non-spin-flip (NSF) data were collected with the neutron polarization oriented vertically, $P_z$, for two different positions of the the 60$^{\circ}$ detector bank to cover the $2\theta$ scattering range 3 - 112$^{\circ}$. In this configuration, the SF setting measures the magnetic scattering, and the NSF gives the combined nuclear and magnetic contributions. The nuclear spin-incoherent scattering contribution from MnTe is expected to be insignificant. A flipping ratio of 13 was determined by evaluating the scattering from nuclear Bragg peaks in SF and NSF configurations.
Post-processing data corrections for the angle-dependent supermirror transmission and for the flipping ratio efficiency were carried out using the algorithms implemented in MANTID software, as discussed in Refs.~\onlinecite{zaliz;jpconfs17,savic;jpconfs17}. The magnetic scattering used in this study corresponds to the SF intensity corrected for the flipping ratio according to Eq.~2 in Ref.~\onlinecite{zaliz;jpconfs17}.

\subsection{\label{subsec:results}Results} 
\subsubsection{\label{scattering}Magnetic Scattering}
\begin{figure}
\includegraphics[width=8.0cm]{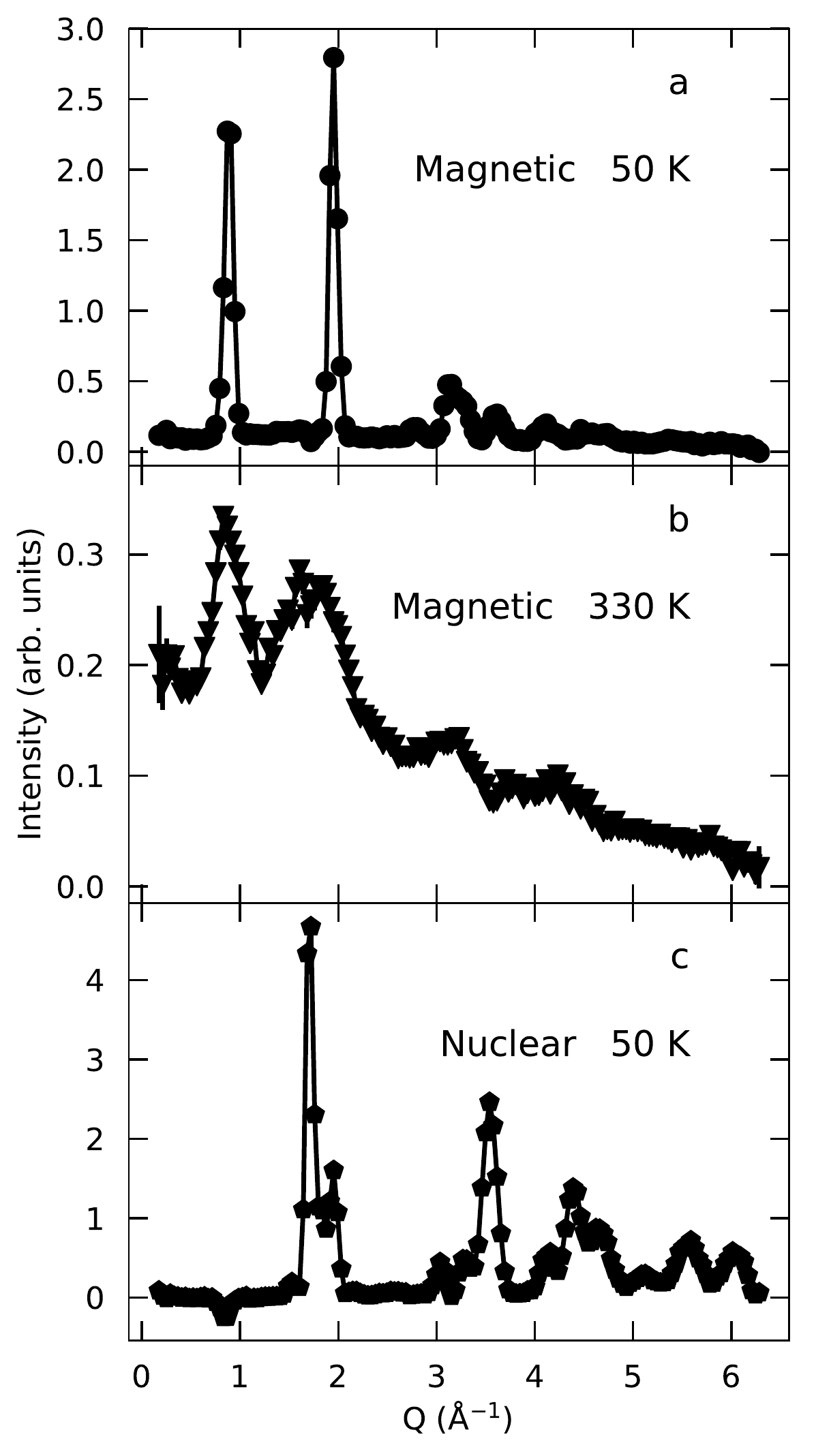}
\caption{Magnetic scattering intensity for MnTe at (a) 50~K and (b) 330~K. Sharp Bragg peaks are present at 50~K due to the long-range ordered antiferromagnetic phase, whereas only diffuse peaks remain at 330~K in the correlated paramagnetic regime. (c) Nuclear scattering at 50~K.}
\label{fig:iq}
\end{figure}

In Fig.~\ref{fig:iq}, we display the magnetic scattering at 50~K (a) and 330~K (b).
Sharp antiferromagnetic Bragg peaks are present at 50~K. At 330~K, the magnetic scattering is diffuse but still quite structured, indicating the persistence of short-range correlations into the paramagnetic phase, as expected from earlier results. The nuclear scattering cross section at 50~K, obtained by subtracting the SF scattering from the NSF scattering, is shown in Fig.~\ref{fig:iq}(c). A slight oversubtraction of the strong magnetic scattering is visible around 0.9~\AA$^{-1}$, but we see minimal interference effects elsewhere, so we conclude that the separation of the different cross sections is sufficiently accurate for our purposes.

\subsubsection{\label{mPDFdata}mPDF Data and Fits}
\begin{figure}
\includegraphics[width=8.0cm]{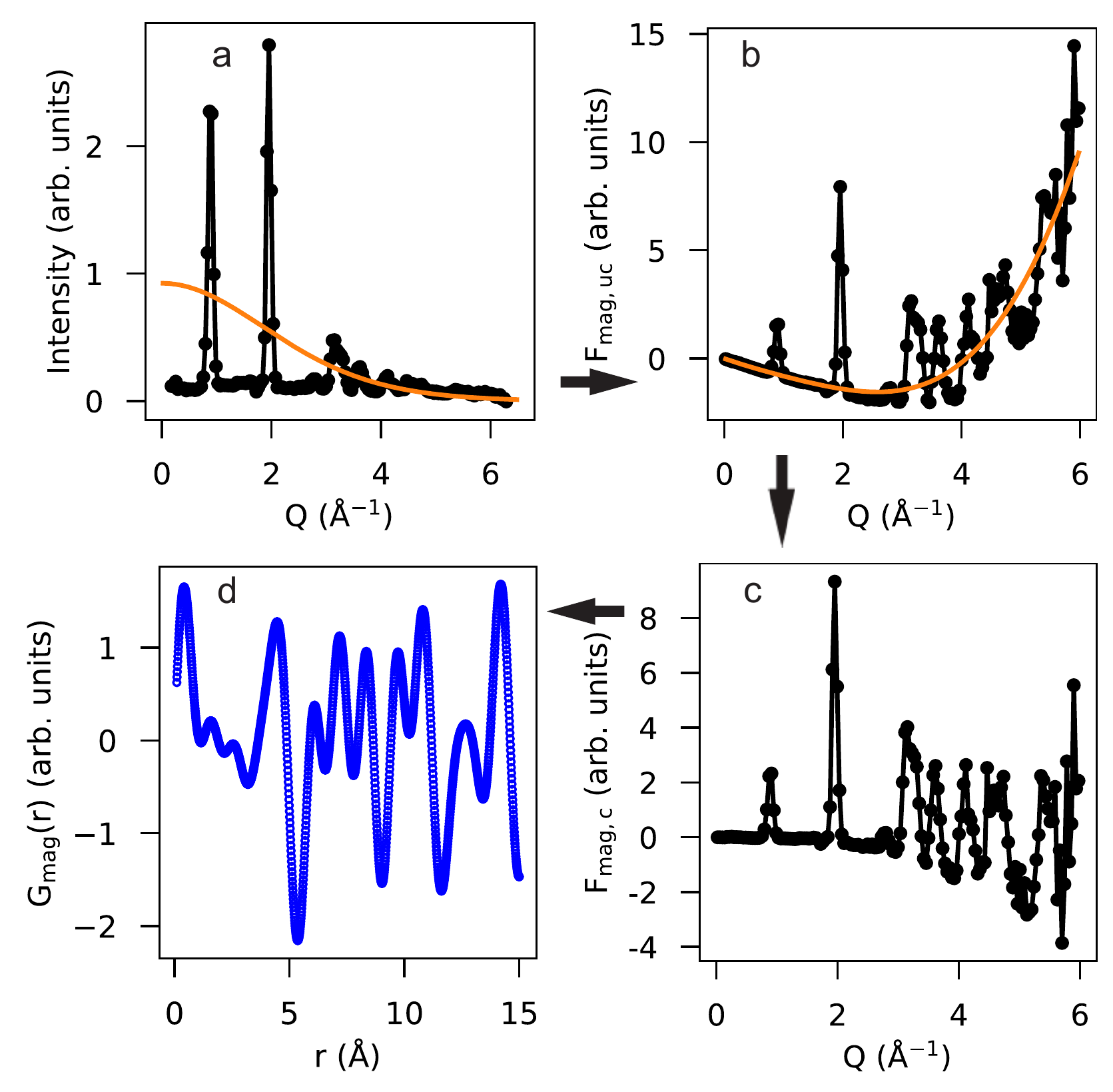}
\caption{Procedure for obtaining the mPDF from magnetic scattering data from MnTe at 50~K. (a) Magnetic scattering intensity (black symbols) overlaid by the square of the magnetic form factor of Mn$^{2+}$ (orange curve; scaled to match the magnitude of the data at high $Q$). (b) Uncorrected reduced magnetic structure function $F_{\mathrm{mag, uc}}$ (black symbols) with the best-fit degree-3 polynomial correction function (orange curve). (c) Corrected $F_{\mathrm{mag, c}}$ after subtracting the polynomial fit. (d) Resulting mPDF using $Q_{\mathrm{max}}=5.5$~\AA$^{-1}$. }
\label{fig:getmPDF}
\end{figure}

Fig.~\ref{fig:getmPDF} shows key results from the \textit{ad hoc} data correction process to produce the mPDF from the scattering data collected at 50~K.
In panel (a), we display the magnetic scattering data and the squared magnetic form factor for Mn$^{2+}$, scaled to match the scattering profile between 3 and 6~\AA$^{-1}$ as closely as possible. We used the analytical approximation of the form factor as reported in the \textit{International Tables for Crystallography}~\cite{wilso;b;itc95}. Panel (b) shows the uncorrected form of the reduced magnetic structure function \Fmag\ (not to be confused with the magnetic structure factor used in conventional magnetic diffraction analysis), given by
\begin{align}
F_{\mathrm{mag, uc}}(Q)&=Q\left(\frac{\left(\text{d}\sigma/\text{d}\Omega\right)_{\mathrm{mag}}}{A [f(Q)]^2}-1\right) \label{Fmag},
\end{align}
where $A$ is the scaling constant used in panel (a), replacing the coefficient on $[f(Q)]^2$ shown in Eq.~\ref{FT} that would be appropriate if the data were on an absolute scale. The ``uc'' in the subscript of the vertical axis label stands for ``uncorrected''. Note that the increasing behavior of \Fmag\ at high $Q$ is non-physical, indicating an imperfect normalization by the squared magnetic form factor. Indeed, the correct form of \Fmag\ should oscillate around zero at high $Q$, since the correct normalization of the magnetic scattering in Eq.~\ref{Fmag} puts it on the scale of unity. To correct this error, we used least-squares minimization to fit a polynomial to $F_{\mathrm{mag, uc}}$, in accordance with the \textit{ad hoc} approach used for x-ray PDF in PDFgetX3~\cite{juhas;jac13}. Fig.~\ref{fig:getmPDF}(b) shows the best-fit cubic polynomial as the orange curve [note that this is not the squared magnetic form factor, which is instead displayed in Fig.~\ref{fig:getmPDF}(a)]. With a polynomial of degree 3 and $Q_{\mathrm{maxinst}}=6.0$~\AA$^{-1}$, the corresponding value of $r_{\mathrm{poly}}$ is 1.57~\AA, safely below the nearest-neighbor Mn-Mn distance of $\sim$3.37~\AA. Due to the relatively low degree of the polynomial, it effectively captures the slowly modulating, nonphysical behavior of $F_{\mathrm{mag, uc}}$ without fitting to the physically meaningful features in the data, which have characteristically faster modulations in $Q$. This polynomial is sometimes referred to as a background function, which is accurate in the sense that it captures the slowly modulating errors in $F_{\mathrm{mag, uc}}$ due to imperfect normalization by the squared magnetic form factor, but it should not be interpreted as an attempt to correct for the background scattering from the instrumental setup. We assume the instrumental background has already been removed. In panel (c), we display the corrected form of \Fmag, labeled $F_{\mathrm{mag, c}}$, resulting from subtracting the fitted polynomial. We note that $F_{\mathrm{mag, c}}$ now oscillates around zero as $Q$ becomes large, as expected when all errors have been removed. Finally, panel (d) shows the resulting mPDF using $Q_{\mathrm{max}}=5.5$~\AA$^{-1}$, $Q_{\mathrm{maxinst}}=6.0$~\AA$^{-1}$, and $r_{\mathrm{poly}} = 1.57$~\AA.

\begin{figure}
\includegraphics[width=8.0cm]{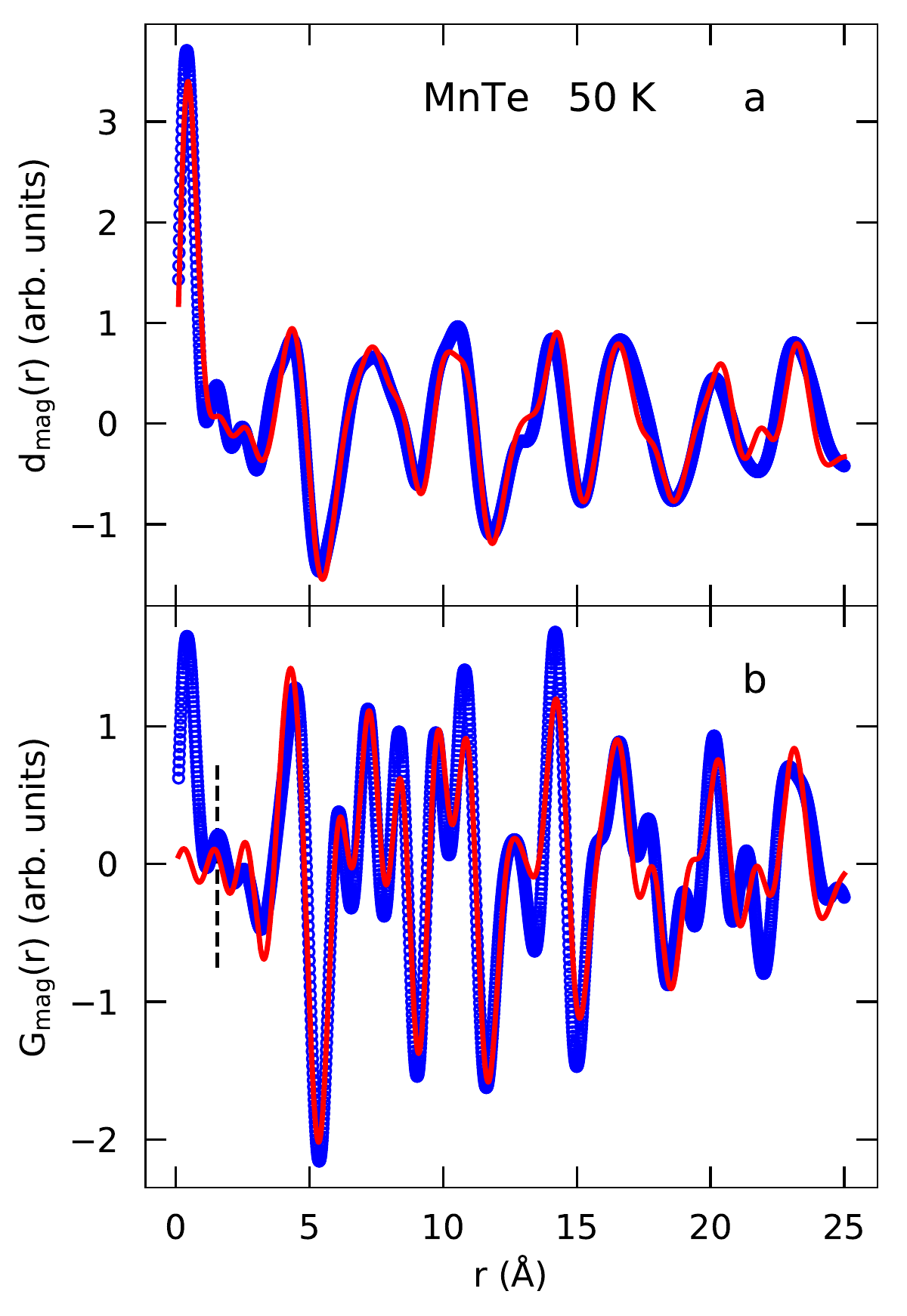}
\caption{Magnetic PDF data for MnTe at 50~K, including (a) the non-deconvoluted mPDF \dmag\ and (b) the proper mPDF \Gmag. The blue circles and red curve show the experimental data and best-fit mPDF, respectively. In panel (b), the vertical dashed line indicates the selected value of $r_{\mathrm{poly}}=1.55$~\AA, below which artifacts from the polynomial correction function are expected to be significant. }
\label{fig:mPDF50K}
\end{figure}

In Fig.~\ref{fig:mPDF50K}, we display fits to the mPDF data at 50~K using the published magnetic structure of MnTe.
We plot the non-deconvoluted mPDF data (given by Eq.~\ref{eq;drexp}) and fit in panel (a) in blue symbols and a solid red curve, respectively. The fit agrees closely with the data and is free from the high-frequency noise that accompanies non-deconvoluted mPDF data obtained together with standard atomic PDF data, highlighting the advantage of using polarized neutrons. In Fig.~\ref{fig:mPDF50K}(b), we plot the deconvoluted mPDF data and fit, where the greatly improved real-space resolution is immediately apparent. The fit successfully captures the most prominent features of the mPDF data, demonstrating the success of our approach. As fitting parameters, we included an arbitrary scale factor, a real-space damping term to account for the finite $Q$-space resolution of the data (known as $Q_{damp}$ in conventional PDF analysis), and an $r$-dependent real-space broadening term ($Q_{broad}$ for conventional PDF) to account for statistical noise in the scattering data~\cite{farro;jpcm07}. The fitted values of $Q_{damp}$ and $Q_{broad}$ were 0.037~\AA$^{-1}$ and 1.68~\AA$^{-1}$, respectively. We note the presence of artifacts for $r \lesssim 1.5$~\AA\ (corresponding closely to $r_{\mathrm{poly}} = 1.57$~\AA) and minor misfits throughout the full data range, but considering that this experimental mPDF curve was generated without any human intervention or fine tuning (in contrast to earlier efforts~\cite{frand;aca15, kodam;jpscp21}), the overall level of agreement is highly encouraging. We further note that, ignoring any distortions or errors in the data, the mPDF pattern in the ordered state is equivalent to the mPDF that would be obtained if one were to perform a perfect nuclear and magnetic Rietveld refinement to an unpolarized neutron diffraction pattern over a sufficiently large $Q$-range and then apply the normalization and Fourier transform procedure to the magnetic part of the Rietveld fit.

\begin{figure}
\includegraphics[width=8.0cm]{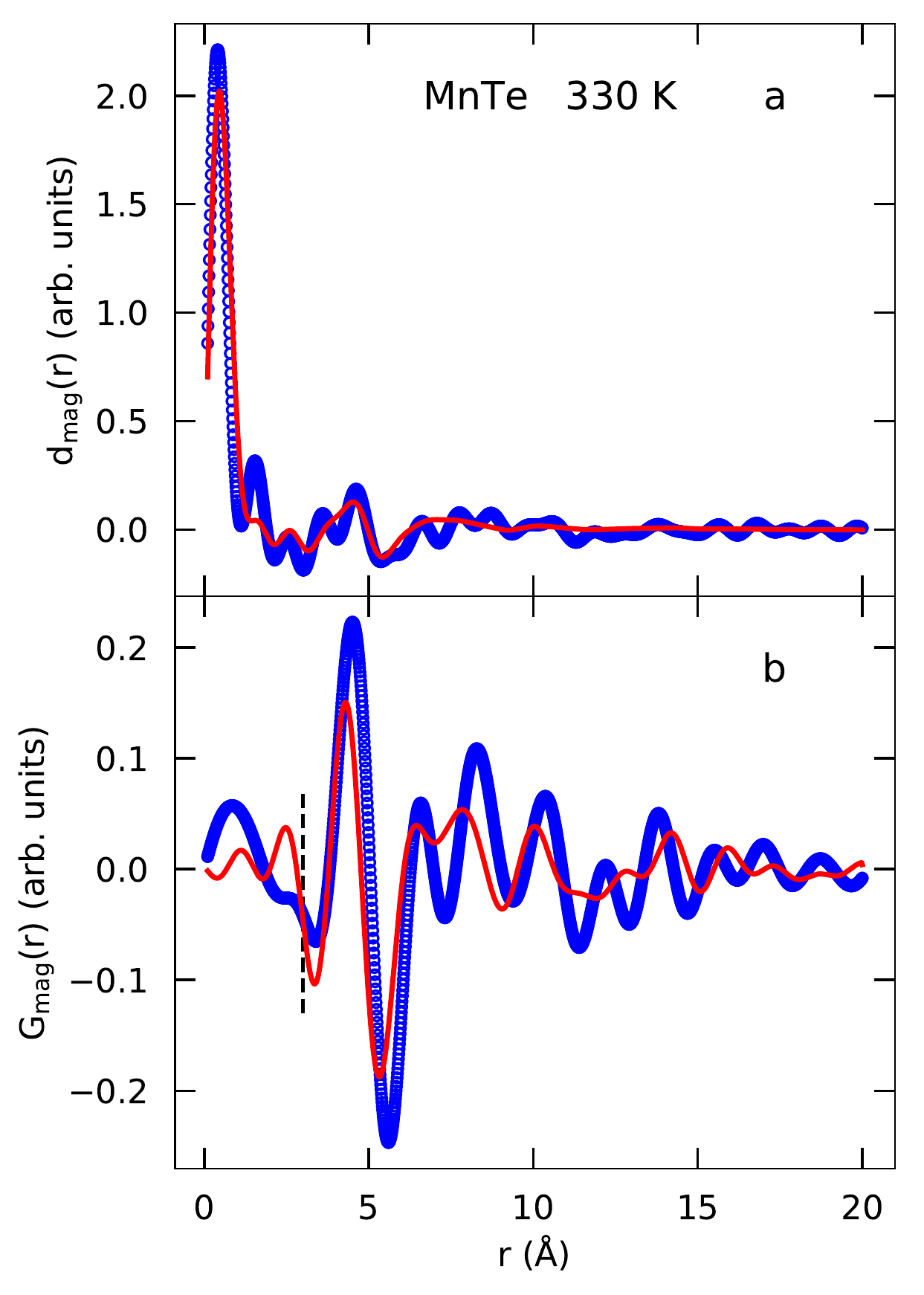}
\caption{Magnetic PDF data for MnTe at 330~K, including (a) the non-deconvoluted mPDF \dmag\ and (b) the proper mPDF \Gmag. The blue circles and red curve show the experimental data and best-fit mPDF with an anisotropic correlation model, respectively. In panel (b), the vertical dashed line indicates the selected value of $r_{\mathrm{poly}}=3.0$~\AA. }
\label{fig:mPDF330K}
\end{figure}

In many cases, short-range magnetic correlations will give rise to the most physically interesting mPDF data. In that spirit, we plot in Fig.~\ref{fig:mPDF330K} the mPDF data and fits for MnTe at 330~K, somewhat above $T_{\mathrm{N}}=307$~K and in the ``correlated paramagnet'' regime where short-range correlations persist in the absence of long-range magnetic order.
Once again, the non-deconvoluted and deconvoluted mPDF patterns are displayed in panels (a) and (b), respectively. Because the magnetic scattering is significantly weaker in the paramagnetic phase, the $Q$-range of usable scattering data for the Fourier transform was more limited, so the data shown were generated with $Q_{\mathrm{max}} = 4.5$~\AA$^{-1}$, together with $r_{\mathrm{poly}}=3.0$~\AA\ and $Q_{\mathrm{maxinst}} = 6.0$~\AA$^{-1}$. We also used a Fermi-Dirac modification function with a width of 1.0~\AA$^{-1}$ to generate the deconvoluted mPDF (see next section for details about the modification function). The non-deconvoluted fit in panel (a) is reasonably good, although truncation effects are apparent in the data. The deconvoluted mPDF data in Fig.~\ref{fig:mPDF330K}(b) exhibit somewhat larger artifacts that the \textit{ad hoc} corrections could not remedy, but the near-neighbor antiferromagnetic correlations are clearly evident in the data, e.g. as the alternating negative and positive peaks around 3.4~\AA\ (nearest neighbor, antiparallel), 4.2~\AA\ (second nearest neighbor, parallel), and 5.4~\AA\ (third nearest neighbor, antiparallel). The mPDF fit captures the general features of the data. We attribute the reduced data quality in the correlated paramagnetic state to the more restricted $Q$-range and greater statistical noise in the scattering data compared to the data collected at 50~K. This underscores the need for high-statistics measurements over a sufficient $Q$-range to produce optimal high-resolution mPDF data.

Despite the more limited data quality, we find it encouraging that the fitted magnetic model produced from the properly deconvoluted mPDF data agrees well with the model produced by the fit to the non-deconvoluted data and to our earlier mPDF fits performed on a different instrument with unpolarized neutrons~\cite{baral;matter22}. In particular, the short-range correlations in MnTe were shown to be spatially anisotropic in our earlier study, with a longer correlation length along the crystallographic $c$ axis than within the hexagonal $ab$ plane. The present fits included two free parameters capturing this anisotropic effect in the magnetic model. The fit naturally converged to correlation lengths of 7.1(5) and 4.6(6)~\AA\ along $c$ and in the $ab$ plane, respectively, in good agreement to those reported in Ref.~\onlinecite{baral;matter22} at a similar temperature. The consistency of these results validates the earlier report and the use of polarized neutrons for obtaining useful mPDF data in a correlated paramagnet.

\subsubsection{\label{windows}Effect of Modification Functions}
\begin{figure}
\includegraphics[width=8.0cm]{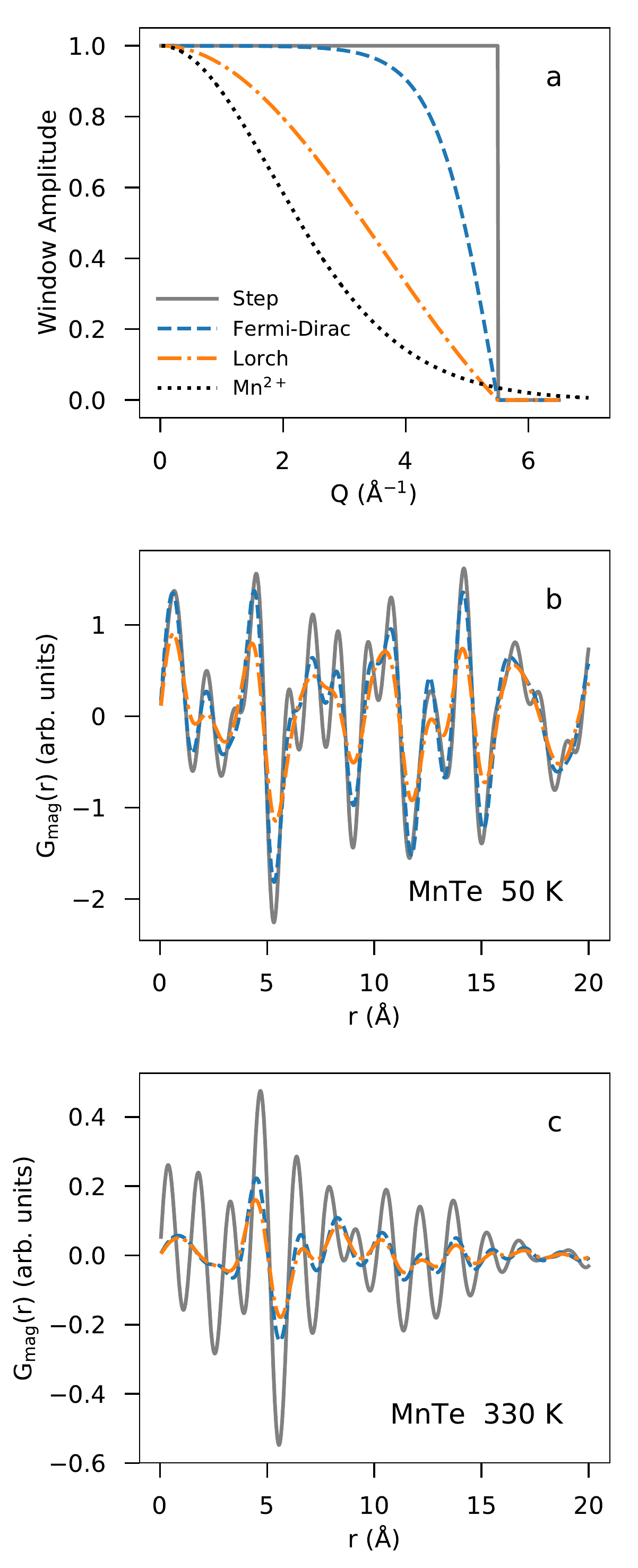}
\caption{(a) modification functions as defined in the main text with $Q_{\mathrm{max}} = 5.5~$\AA$^{-1}$. The Fermi-Dirac modification function uses $\Delta=0.5$~\AA$^{-1}$. The squared magnetic form factor of Mn$^{2+}$ is also shown for reference. (b) The mPDF patterns for MnTe at 50~K resulting from application of the modification functions in (a), using the same legend. (c) Same as (b), but for 330~K. The transform and mask parameters are $Q_{\mathrm{max}} = 4.5~$\AA$^{-1}$ and $\Delta = 1.0~$\AA$^{-1}$ .}
\label{fig:windows}
\end{figure}

Finally, we discuss the effect of applying a $Q$-space modification function~\cite{soper;jac12}, also known as a window function, to $F_{\mathrm{mag, c}}(Q)$ prior to the Fourier transform. In the context of atomic PDF analysis, various modification functions have been used to suppress the effect of statistical or systematic errors in the scattering data at high $Q$, reducing high-frequency ripples in the PDF (but with the cost of reduced real-space resolution and possible loss of physically meaningful information)~\cite{soper;jac12}. A modification function $w(Q)$ operates simply through multiplication with the corrected form of \Fmag, such that $w(Q) \times F_{\mathrm{mag}}(Q)$ is the quantity to be Fourier transformed. We consider here three modification functions: the default step function
\begin{align} 
w_{\mathrm{s}}(Q)=\begin{cases} 1 & \text{if } Q \le Q_{\mathrm{max}} \\ 0 & \text{if } Q > Q_{\mathrm{max}} \end{cases},
\end{align} 
a modified Fermi-Dirac function
\begin{align} 
w_{\mathrm{FD}}(Q)=\begin{cases} \frac{2}{e^{(Q - Q_{\mathrm{max}})/\Delta}+1} - 1 & \text{if } Q \le Q_{\mathrm{max}} \\ 0 & \text{if } Q > Q_{\mathrm{max}} \end{cases},
\end{align} 
and the conventional Lorch function~\cite{lorch;jpcss69}
\begin{align} 
w_{\mathrm{L}}(Q)=\begin{cases} \frac{Q_{\mathrm{max}}}{\pi Q} \sin\left( \frac{\pi Q}{Q_{\mathrm{max}}} \right) & \text{if } Q \le Q_{\mathrm{max}} \\ 0 & \text{if } Q > Q_{\mathrm{max}} \end{cases}.
\end{align} 
We note that $w_{\mathrm{s}}$ is equivalent to applying no modification function at all, since it is unity up until the hard cutoff of the Fourier transform at $Q_{\mathrm{max}}$. These three modification functions are displayed in Fig.~\ref{fig:windows}(a), together with the squared magnetic form factor for Mn$^{2+}$ for reference.
In all cases, we use $Q_{\mathrm{max}} = 5.5~$\AA$^{-1}$, and for $w_{\mathrm{FD}}$, we set $\Delta$ to 0.5~\AA$^{-1}$. The mPDF curves at 50~K generated with the application of each of these modification functions are given in Fig.~\ref{fig:windows}(b). Both $w_{\mathrm{FD}}$ and $w_{\mathrm{L}}$ suppress high-frequency features in the mPDF relative to the default mPDF using $w_{\mathrm{s}}$. The effect is particularly pronounced for $w_{\mathrm{L}}$. However, we note that for the available $Q$-range of 5.5~\AA$^{-1}$ included in the Fourier transform, $w_{\mathrm{L}}$ results in a significant suppression of \Fmag\ over nearly the entire range similar in scale to the squared magnetic form factor shown in Fig.~\ref{fig:windows}(a). In effect, then, application of $w_{\mathrm{L}}$ undoes the normalization by the squared form factor, yielding a quantity that is similar in real-space resolution to the non-deconvoluted mPDF that we were originally hoping to improve upon. We therefore suggest that a Lorch window is of limited usefulness for producing deconvoluted mPDF patterns when the available $Q$-range does not significantly exceed the extent of the squared magnetic form factor. On the other hand, $w_{\mathrm{FD}}$ with a modestly chosen value of $\Delta$ may provide value by suppressing artifacts from high-$Q$ noise without sacrificing meaningful mPDF data due to excessive broadening.

Fig.~\ref{fig:windows}(c) shows the mPDF curves at 330~K obtained for each type of modification function, this time with $Q_{\mathrm{max}} = 4.5~$\AA$^{-1}$ and $\Delta = 1.0$~\AA$^{-1}$. The Lorch and Fermi-Dirac functions again result in greatly suppressed Fourier ripples, with the Fermi-Dirac function preserving somewhat more real-space resolution than the Lorch function.

In the mPDF fits presented in the previous section, we found no need to use a modification function for the data collected at 50~K where the magnetic scattering is strong. However, given the weaker magnetic scattering at 330~K, we found it beneficial to apply a Fermi-Dirac modification function with a width of 1.0~\AA$^{-1}$. We chose this width because, after testing several different values, we found that 1.0~\AA$^{-1}$ offered a good balance between minimizing artifacts in the mPDF data and preserving the real-space resolution to a reasonable degree. Ideally, the calculated mPDF used for modeling the data should be convoluted with the Fourier transform of the modification function to account correctly for the change in real-space resolution caused by the modification function.

\section{\label{sec:discussion}Discussion and Conclusion}
We have reported the first use of polarized neutrons to generate mPDF data and the first application of PDFgetX3-style \textit{ad hoc} corrections to produce high-resolution, properly deconvoluted \Gmag\ curves with minimal human input. Polarization analysis allowed us to isolate the magnetic scattering cross section and produce the resulting mPDF data with little to no interference from nuclear scattering. On HYSPEC, we achieved a $Q_{\mathrm{max}}$ value of 5.5~\AA$^{-1}$ in the magnetically ordered state of MnTe. This was sufficient for an exceptionally clean non-deconvoluted mPDF curve and a high-quality proper mPDF curve produced with the help of the \textit{ad hoc} corrections, demonstrating the promise of this method for producing mPDF data. In the paramagnetic state, we obtained good-quality non-deconvoluted and deconvoluted mPDF curves suitable for quantitative mPDF refinements that confirmed a prior result. However, the weaker magnetic scattering at 330~K necessitated a more limited $Q$-range and introduced significantly more noise and artifacts into the data. These issues could be ameliorated in large part simply by collecting more data to improve the statistics over a large enough range of momentum transfer (ideally up to 5.5~\AA$^{-1}$ or greater). Finally, our results suggest that if a modification function is to be used to reduce high-frequency noise in the mPDF data, a modified Fermi Dirac function or similar is preferred over the traditional Lorch function.

Overall, the results reported here represent a significant step forward in the development of mPDF methodologies by highlighting the promise of polarized neutrons and modern data processing techniques to produce high-quality, high-resoultion mPDF data. We have provided a realistic view of both the capabilities and the limitations posed by current experimental resources when generating high-resolution mPDF data from weak and diffuse magnetic scattering. With major developments in neutron capabilities on the near horizon at the European Spallation Source, the Second Target Station (STS) of the Spallation Neutron Source, and elsewhere, instrumental limitations are expected to be less of a concern in the future. We mention particularly the planned STS instrument VERDI~\cite{garle;rsi22}, for which the combination of high beam intensity, large $Q$ coverage, and polarization analysis should enable routine collection of high-resolution mPDF data of excellent quality, which we expect to transform the landscape of mPDF studies in condensed matter and materials physics.

\section{Supplementary Material}
The supplemental materials include the magnetic scattering cross section at 50~K and 330~K, the nuclear scattering cross section at 50~K and 330~K, python code for generating the mPDF with the \textit{ad hoc} corrections, and an example python script showing how to use the code.

\begin{acknowledgments}
We thank Melissa Graves-Brook for assistance with the HYSPEC experiment. Work by B.A.F. and R.B. was supported by the U.S. Department of Energy, Office of Science, Office of Basic Energy Sciences (DOE-BES) through Award No. DE-SC0021134. This study used resources at the Spallation Neutron Source (SNS), a DOE Office of Science User Facility operated by the Oak Ridge National Laboratory.
\end{acknowledgments}

\section*{Data Availability Statement}

The data that support the findings of this study are available within the article and its supplementary material.

\section*{Author Declarations}

The authors have no conflicts to disclose.


\begin{thebibliography}{34}%
	\makeatletter
	\providecommand \@ifxundefined [1]{%
		\@ifx{#1\undefined}
	}%
	\providecommand \@ifnum [1]{%
		\ifnum #1\expandafter \@firstoftwo
		\else \expandafter \@secondoftwo
		\fi
	}%
	\providecommand \@ifx [1]{%
		\ifx #1\expandafter \@firstoftwo
		\else \expandafter \@secondoftwo
		\fi
	}%
	\providecommand \natexlab [1]{#1}%
	\providecommand \enquote  [1]{``#1''}%
	\providecommand \bibnamefont  [1]{#1}%
	\providecommand \bibfnamefont [1]{#1}%
	\providecommand \citenamefont [1]{#1}%
	\providecommand \href@noop [0]{\@secondoftwo}%
	\providecommand \href [0]{\begingroup \@sanitize@url \@href}%
	\providecommand \@href[1]{\@@startlink{#1}\@@href}%
	\providecommand \@@href[1]{\endgroup#1\@@endlink}%
	\providecommand \@sanitize@url [0]{\catcode `\\12\catcode `\$12\catcode
		`\&12\catcode `\#12\catcode `\^12\catcode `\_12\catcode `\%12\relax}%
	\providecommand \@@startlink[1]{}%
	\providecommand \@@endlink[0]{}%
	\providecommand \url  [0]{\begingroup\@sanitize@url \@url }%
	\providecommand \@url [1]{\endgroup\@href {#1}{\urlprefix }}%
	\providecommand \urlprefix  [0]{URL }%
	\providecommand \Eprint [0]{\href }%
	\providecommand \doibase [0]{http://dx.doi.org/}%
	\providecommand \selectlanguage [0]{\@gobble}%
	\providecommand \bibinfo  [0]{\@secondoftwo}%
	\providecommand \bibfield  [0]{\@secondoftwo}%
	\providecommand \translation [1]{[#1]}%
	\providecommand \BibitemOpen [0]{}%
	\providecommand \bibitemStop [0]{}%
	\providecommand \bibitemNoStop [0]{.\EOS\space}%
	\providecommand \EOS [0]{\spacefactor3000\relax}%
	\providecommand \BibitemShut  [1]{\csname bibitem#1\endcsname}%
	\let\auto@bib@innerbib\@empty
	%</preamble>
	\bibitem [{\citenamefont {Frandsen}, \citenamefont {Yang},\ and\ \citenamefont
		{Billinge}(2014)}]{frand;aca14}%
	\BibitemOpen
	\bibfield  {author} {\bibinfo {author} {\bibfnamefont {B.~A.}\ \bibnamefont
			{Frandsen}}, \bibinfo {author} {\bibfnamefont {X.}~\bibnamefont {Yang}}, \
		and\ \bibinfo {author} {\bibfnamefont {S.~J.~L.}\ \bibnamefont {Billinge}},\
	}\bibfield  {title} {\enquote {\bibinfo {title} {Magnetic pair distribution
				function analysis of local magnetic correlations},}\ }\href {\doibase
		10.1107/S2053273313033081} {\bibfield  {journal} {\bibinfo  {journal} {Acta
				Crystallogr. A}\ }\textbf {\bibinfo {volume} {70}},\ \bibinfo {pages} {3--11}
		(\bibinfo {year} {2014})}\BibitemShut {NoStop}%
	\bibitem [{\citenamefont {Frandsen}\ and\ \citenamefont
		{Billinge}(2015)}]{frand;aca15}%
	\BibitemOpen
	\bibfield  {author} {\bibinfo {author} {\bibfnamefont {B.~A.}\ \bibnamefont
			{Frandsen}}\ and\ \bibinfo {author} {\bibfnamefont {S.~J.~L.}\ \bibnamefont
			{Billinge}},\ }\bibfield  {title} {\enquote {\bibinfo {title} {Magnetic
				structure determination from the magnetic pair distribution function
				{(mPDF)}: ground state of {MnO}},}\ }\href {\doibase
		10.1107/S205327331500306X} {\bibfield  {journal} {\bibinfo  {journal} {Acta
				Crystallogr. A}\ }\textbf {\bibinfo {volume} {71}},\ \bibinfo {pages}
		{325--334} (\bibinfo {year} {2015})}\BibitemShut {NoStop}%
	\bibitem [{\citenamefont {Egami}\ and\ \citenamefont
		{Billinge}(2012)}]{egami;b;utbp12}%
	\BibitemOpen
	\bibfield  {author} {\bibinfo {author} {\bibfnamefont {T.}~\bibnamefont
			{Egami}}\ and\ \bibinfo {author} {\bibfnamefont {S.~J.~L.}\ \bibnamefont
			{Billinge}},\ }\href
	{http://store.elsevier.com/product.jsp?lid=0\&iid=73\&sid=0\&isbn=9780080971414}
	{\emph {\bibinfo {title} {Underneath the Bragg peaks: structural analysis of
				complex materials}}},\ \bibinfo {edition} {2nd}\ ed.\ (\bibinfo  {publisher}
	{Elsevier},\ \bibinfo {address} {Amsterdam},\ \bibinfo {year}
	{2012})\BibitemShut {NoStop}%
	\bibitem [{\citenamefont {Frandsen}\ \emph
		{et~al.}(2016{\natexlab{a}})\citenamefont {Frandsen}, \citenamefont
		{Brunelli}, \citenamefont {Page}, \citenamefont {Uemura}, \citenamefont
		{Staunton},\ and\ \citenamefont {Billinge}}]{frand;prl16}%
	\BibitemOpen
	\bibfield  {author} {\bibinfo {author} {\bibfnamefont {B.~A.}\ \bibnamefont
			{Frandsen}}, \bibinfo {author} {\bibfnamefont {M.}~\bibnamefont {Brunelli}},
		\bibinfo {author} {\bibfnamefont {K.}~\bibnamefont {Page}}, \bibinfo {author}
		{\bibfnamefont {Y.~J.}\ \bibnamefont {Uemura}}, \bibinfo {author}
		{\bibfnamefont {J.~B.}\ \bibnamefont {Staunton}}, \ and\ \bibinfo {author}
		{\bibfnamefont {S.~J.~L.}\ \bibnamefont {Billinge}},\ }\bibfield  {title}
	{\enquote {\bibinfo {title} {Verification of anderson superexchange in mno
				via magnetic pair distribution function analysis and \textit{ab initio}
				theory},}\ }\href {\doibase 10.1103/PhysRevLett.116.197204} {\bibfield
		{journal} {\bibinfo  {journal} {Phys. Rev. Lett.}\ }\textbf {\bibinfo
			{volume} {116}},\ \bibinfo {pages} {197204} (\bibinfo {year}
		{2016}{\natexlab{a}})}\BibitemShut {NoStop}%
	\bibitem [{\citenamefont {Frandsen}\ \emph
		{et~al.}(2016{\natexlab{b}})\citenamefont {Frandsen}, \citenamefont {Gong},
		\citenamefont {Terban}, \citenamefont {Banerjee}, \citenamefont {Chen},
		\citenamefont {Jin}, \citenamefont {Feygenson}, \citenamefont {Uemura},\ and\
		\citenamefont {Billinge}}]{frand;prb16}%
	\BibitemOpen
	\bibfield  {author} {\bibinfo {author} {\bibfnamefont {B.~A.}\ \bibnamefont
			{Frandsen}}, \bibinfo {author} {\bibfnamefont {Z.}~\bibnamefont {Gong}},
		\bibinfo {author} {\bibfnamefont {M.~W.}\ \bibnamefont {Terban}}, \bibinfo
		{author} {\bibfnamefont {S.}~\bibnamefont {Banerjee}}, \bibinfo {author}
		{\bibfnamefont {B.}~\bibnamefont {Chen}}, \bibinfo {author} {\bibfnamefont
			{C.}~\bibnamefont {Jin}}, \bibinfo {author} {\bibfnamefont {M.}~\bibnamefont
			{Feygenson}}, \bibinfo {author} {\bibfnamefont {Y.~J.}\ \bibnamefont
			{Uemura}}, \ and\ \bibinfo {author} {\bibfnamefont {S.~J.~L.}\ \bibnamefont
			{Billinge}},\ }\bibfield  {title} {\enquote {\bibinfo {title} {Local atomic
				and magnetic structure of dilute magnetic semiconductor
				{(Ba,K)(Zn,Mn)$_2$As$_2$}},}\ }\href {\doibase 10.1103/PhysRevB.94.094102}
	{\bibfield  {journal} {\bibinfo  {journal} {Phys. Rev. B}\ }\textbf {\bibinfo
			{volume} {94}},\ \bibinfo {pages} {094102} (\bibinfo {year}
		{2016}{\natexlab{b}})},\ \bibinfo {note} {selected as Editors'
		Suggestion}\BibitemShut {NoStop}%
	\bibitem [{\citenamefont {Frandsen}\ \emph {et~al.}(2017)\citenamefont
		{Frandsen}, \citenamefont {Ross}, \citenamefont {Krizan}, \citenamefont
		{Nilsen}, \citenamefont {Wildes}, \citenamefont {Cava}, \citenamefont
		{Birgeneau},\ and\ \citenamefont {Billinge}}]{frand;prm17}%
	\BibitemOpen
	\bibfield  {author} {\bibinfo {author} {\bibfnamefont {B.~A.}\ \bibnamefont
			{Frandsen}}, \bibinfo {author} {\bibfnamefont {K.~A.}\ \bibnamefont {Ross}},
		\bibinfo {author} {\bibfnamefont {J.~W.}\ \bibnamefont {Krizan}}, \bibinfo
		{author} {\bibfnamefont {G.~J.}\ \bibnamefont {Nilsen}}, \bibinfo {author}
		{\bibfnamefont {A.~R.}\ \bibnamefont {Wildes}}, \bibinfo {author}
		{\bibfnamefont {R.~J.}\ \bibnamefont {Cava}}, \bibinfo {author}
		{\bibfnamefont {R.~J.}\ \bibnamefont {Birgeneau}}, \ and\ \bibinfo {author}
		{\bibfnamefont {S.~J.~L.}\ \bibnamefont {Billinge}},\ }\bibfield  {title}
	{\enquote {\bibinfo {title} {Real-space investigation of short-range magnetic
				correlations in fluoride pyrochlores ${\mathrm{nacaco}}_{2}{\mathrm{f}}_{7}$
				and ${\mathrm{nasrco}}_{2}{\mathrm{f}}_{7}$ with magnetic pair distribution
				function analysis},}\ }\href {\doibase 10.1103/PhysRevMaterials.1.074412}
	{\bibfield  {journal} {\bibinfo  {journal} {Phys. Rev. Materials}\ }\textbf
		{\bibinfo {volume} {1}},\ \bibinfo {pages} {074412} (\bibinfo {year}
		{2017})}\BibitemShut {NoStop}%
	\bibitem [{\citenamefont {Kodama}\ \emph {et~al.}(2017)\citenamefont {Kodama},
		\citenamefont {Ikeda}, \citenamefont {Shamoto},\ and\ \citenamefont
		{Otomo}}]{kodam;jpsj17}%
	\BibitemOpen
	\bibfield  {author} {\bibinfo {author} {\bibfnamefont {K.}~\bibnamefont
			{Kodama}}, \bibinfo {author} {\bibfnamefont {K.}~\bibnamefont {Ikeda}},
		\bibinfo {author} {\bibfnamefont {S.-i.}\ \bibnamefont {Shamoto}}, \ and\
		\bibinfo {author} {\bibfnamefont {T.}~\bibnamefont {Otomo}},\ }\bibfield
	{title} {\enquote {\bibinfo {title} {{Alternative Equation on Magnetic Pair
					Distribution Function for Quantitative Analysis}},}\ }\href {\doibase
		10.7566/JPSJ.86.124708} {\bibfield  {journal} {\bibinfo  {journal} {J. Phys.
				Soc. Jpn}\ }\textbf {\bibinfo {volume} {86}},\ \bibinfo {pages} {124708}
		(\bibinfo {year} {2017})}\BibitemShut {NoStop}%
	\bibitem [{\citenamefont {Roth}\ \emph {et~al.}(2019)\citenamefont {Roth},
		\citenamefont {Ye}, \citenamefont {May}, \citenamefont {Chakoumakos},\ and\
		\citenamefont {Iversen}}]{roth;prb19}%
	\BibitemOpen
	\bibfield  {author} {\bibinfo {author} {\bibfnamefont {N.}~\bibnamefont
			{Roth}}, \bibinfo {author} {\bibfnamefont {F.}~\bibnamefont {Ye}}, \bibinfo
		{author} {\bibfnamefont {A.~F.}\ \bibnamefont {May}}, \bibinfo {author}
		{\bibfnamefont {B.~C.}\ \bibnamefont {Chakoumakos}}, \ and\ \bibinfo {author}
		{\bibfnamefont {B.~B.}\ \bibnamefont {Iversen}},\ }\bibfield  {title}
	{\enquote {\bibinfo {title} {{Magnetic correlations and structure in bixbyite
					across the spin-glass transition}},}\ }\href {\doibase
		10.1103/PhysRevB.100.144404} {\bibfield  {journal} {\bibinfo  {journal}
			{Phys. Rev. B}\ }\textbf {\bibinfo {volume} {100}},\ \bibinfo {pages}
		{144404} (\bibinfo {year} {2019})}\BibitemShut {NoStop}%
	\bibitem [{\citenamefont {Tripathi}\ \emph {et~al.}(2019)\citenamefont
		{Tripathi}, \citenamefont {Chatterji}, \citenamefont {Fischer}, \citenamefont
		{Raghunathan}, \citenamefont {Majumder}, \citenamefont {Choudhary},\ and\
		\citenamefont {Phase}}]{tripa;prb19}%
	\BibitemOpen
	\bibfield  {author} {\bibinfo {author} {\bibfnamefont {M.}~\bibnamefont
			{Tripathi}}, \bibinfo {author} {\bibfnamefont {T.}~\bibnamefont {Chatterji}},
		\bibinfo {author} {\bibfnamefont {H.~E.}\ \bibnamefont {Fischer}}, \bibinfo
		{author} {\bibfnamefont {R.}~\bibnamefont {Raghunathan}}, \bibinfo {author}
		{\bibfnamefont {S.}~\bibnamefont {Majumder}}, \bibinfo {author}
		{\bibfnamefont {R.~J.}\ \bibnamefont {Choudhary}}, \ and\ \bibinfo {author}
		{\bibfnamefont {D.~M.}\ \bibnamefont {Phase}},\ }\bibfield  {title} {\enquote
		{\bibinfo {title} {Role of local short-scale correlations in the mechanism of
				negative magnetization},}\ }\href {\doibase 10.1103/PhysRevB.99.014422}
	{\bibfield  {journal} {\bibinfo  {journal} {Phys. Rev. B}\ }\textbf {\bibinfo
			{volume} {99}},\ \bibinfo {pages} {014422} (\bibinfo {year}
		{2019})}\BibitemShut {NoStop}%
	\bibitem [{\citenamefont {Lefran\c{c}ois}\ \emph {et~al.}(2019)\citenamefont
		{Lefran\c{c}ois}, \citenamefont {Mangin-Thro}, \citenamefont {Lhotel},
		\citenamefont {Robert}, \citenamefont {Petit}, \citenamefont {Cathelin},
		\citenamefont {Fischer}, \citenamefont {Colin}, \citenamefont {Damay},
		\citenamefont {Ollivier}, \citenamefont {Lejay}, \citenamefont {Chapon},
		\citenamefont {Simonet},\ and\ \citenamefont {Ballou}}]{lefra;prb19}%
	\BibitemOpen
	\bibfield  {author} {\bibinfo {author} {\bibfnamefont {E.}~\bibnamefont
			{Lefran\c{c}ois}}, \bibinfo {author} {\bibfnamefont {L.}~\bibnamefont
			{Mangin-Thro}}, \bibinfo {author} {\bibfnamefont {E.}~\bibnamefont {Lhotel}},
		\bibinfo {author} {\bibfnamefont {J.}~\bibnamefont {Robert}}, \bibinfo
		{author} {\bibfnamefont {S.}~\bibnamefont {Petit}}, \bibinfo {author}
		{\bibfnamefont {V.}~\bibnamefont {Cathelin}}, \bibinfo {author}
		{\bibfnamefont {H.~E.}\ \bibnamefont {Fischer}}, \bibinfo {author}
		{\bibfnamefont {C.~V.}\ \bibnamefont {Colin}}, \bibinfo {author}
		{\bibfnamefont {F.}~\bibnamefont {Damay}}, \bibinfo {author} {\bibfnamefont
			{J.}~\bibnamefont {Ollivier}}, \bibinfo {author} {\bibfnamefont
			{P.}~\bibnamefont {Lejay}}, \bibinfo {author} {\bibfnamefont {L.~C.}\
			\bibnamefont {Chapon}}, \bibinfo {author} {\bibfnamefont {V.}~\bibnamefont
			{Simonet}}, \ and\ \bibinfo {author} {\bibfnamefont {R.}~\bibnamefont
			{Ballou}},\ }\bibfield  {title} {\enquote {\bibinfo {title} {{Spin decoupling
					under a staggered field in the Gd$_2$Ir$_2$O$_7$ pyrochlore}},}\ }\href
	{\doibase 10.1103/PhysRevB.99.060401} {\bibfield  {journal} {\bibinfo
			{journal} {Phys. Rev. B}\ }\textbf {\bibinfo {volume} {99}},\ \bibinfo
		{pages} {060401} (\bibinfo {year} {2019})}\BibitemShut {NoStop}%
	\bibitem [{\citenamefont {Zhang}\ \emph {et~al.}(2019)\citenamefont {Zhang},
		\citenamefont {Scholz}, \citenamefont {Dronskowski}, \citenamefont
		{McDonnell},\ and\ \citenamefont {Tucker}}]{zhang;prb19}%
	\BibitemOpen
	\bibfield  {author} {\bibinfo {author} {\bibfnamefont {Y.}~\bibnamefont
			{Zhang}}, \bibinfo {author} {\bibfnamefont {T.}~\bibnamefont {Scholz}},
		\bibinfo {author} {\bibfnamefont {R.}~\bibnamefont {Dronskowski}}, \bibinfo
		{author} {\bibfnamefont {M.~T.}\ \bibnamefont {McDonnell}}, \ and\ \bibinfo
		{author} {\bibfnamefont {M.~G.}\ \bibnamefont {Tucker}},\ }\bibfield  {title}
	{\enquote {\bibinfo {title} {{Local magnetic cluster size identified by
					neutron total scattering in the site-diluted spin glass Sn$_x$Fe$_{4-x}$N
					($x$ = 0.88)}},}\ }\href {\doibase 10.1103/PhysRevB.100.014419} {\bibfield
		{journal} {\bibinfo  {journal} {Phys. Rev. B}\ }\textbf {\bibinfo {volume}
			{100}},\ \bibinfo {pages} {014419} (\bibinfo {year} {2019})}\BibitemShut
	{NoStop}%
	\bibitem [{\citenamefont {Frandsen}\ \emph {et~al.}(2020)\citenamefont
		{Frandsen}, \citenamefont {Bozin}, \citenamefont {Aza}, \citenamefont
		{Mart\'{\i}nez}, \citenamefont {Feygenson}, \citenamefont {Page},\ and\
		\citenamefont {Lappas}}]{frand;prb20}%
	\BibitemOpen
	\bibfield  {author} {\bibinfo {author} {\bibfnamefont {B.~A.}\ \bibnamefont
			{Frandsen}}, \bibinfo {author} {\bibfnamefont {E.~S.}\ \bibnamefont {Bozin}},
		\bibinfo {author} {\bibfnamefont {E.}~\bibnamefont {Aza}}, \bibinfo {author}
		{\bibfnamefont {A.~F.}\ \bibnamefont {Mart\'{\i}nez}}, \bibinfo {author}
		{\bibfnamefont {M.}~\bibnamefont {Feygenson}}, \bibinfo {author}
		{\bibfnamefont {K.}~\bibnamefont {Page}}, \ and\ \bibinfo {author}
		{\bibfnamefont {A.}~\bibnamefont {Lappas}},\ }\bibfield  {title} {\enquote
		{\bibinfo {title} {Nanoscale degeneracy lifting in a geometrically frustrated
				antiferromagnet},}\ }\href {\doibase 10.1103/PhysRevB.101.024423} {\bibfield
		{journal} {\bibinfo  {journal} {Phys. Rev. B}\ }\textbf {\bibinfo {volume}
			{101}},\ \bibinfo {pages} {024423} (\bibinfo {year} {2020})}\BibitemShut
	{NoStop}%
	\bibitem [{\citenamefont {Dun}\ \emph {et~al.}(2021)\citenamefont {Dun},
		\citenamefont {Daum}, \citenamefont {Baral}, \citenamefont {Fischer},
		\citenamefont {Cao}, \citenamefont {Liu}, \citenamefont {Stone},
		\citenamefont {Rodriguez-Rivera}, \citenamefont {Choi}, \citenamefont
		{Huang}, \citenamefont {Zhou}, \citenamefont {Mourigal},\ and\ \citenamefont
		{Frandsen}}]{frand;prb21}%
	\BibitemOpen
	\bibfield  {author} {\bibinfo {author} {\bibfnamefont {Z.}~\bibnamefont
			{Dun}}, \bibinfo {author} {\bibfnamefont {M.}~\bibnamefont {Daum}}, \bibinfo
		{author} {\bibfnamefont {R.}~\bibnamefont {Baral}}, \bibinfo {author}
		{\bibfnamefont {H.~E.}\ \bibnamefont {Fischer}}, \bibinfo {author}
		{\bibfnamefont {H.}~\bibnamefont {Cao}}, \bibinfo {author} {\bibfnamefont
			{Y.}~\bibnamefont {Liu}}, \bibinfo {author} {\bibfnamefont {M.~B.}\
			\bibnamefont {Stone}}, \bibinfo {author} {\bibfnamefont {J.~A.}\ \bibnamefont
			{Rodriguez-Rivera}}, \bibinfo {author} {\bibfnamefont {E.~S.}\ \bibnamefont
			{Choi}}, \bibinfo {author} {\bibfnamefont {Q.}~\bibnamefont {Huang}},
		\bibinfo {author} {\bibfnamefont {H.}~\bibnamefont {Zhou}}, \bibinfo {author}
		{\bibfnamefont {M.}~\bibnamefont {Mourigal}}, \ and\ \bibinfo {author}
		{\bibfnamefont {B.~A.}\ \bibnamefont {Frandsen}},\ }\bibfield  {title}
	{\enquote {\bibinfo {title} {{Neutron scattering investigation of proposed
					Kosterlitz-Thouless transitions in the triangular-lattice Ising
					antiferromagnet ${\mathrm{TmMgGaO}}_{4}$}},}\ }\href {\doibase
		10.1103/PhysRevB.103.064424} {\bibfield  {journal} {\bibinfo  {journal}
			{Phys. Rev. B}\ }\textbf {\bibinfo {volume} {103}},\ \bibinfo {pages}
		{064424} (\bibinfo {year} {2021})}\BibitemShut {NoStop}%
	\bibitem [{\citenamefont {Baral}\ \emph {et~al.}(2022)\citenamefont {Baral},
		\citenamefont {Christensen}, \citenamefont {Hamilton}, \citenamefont {Ye},
		\citenamefont {Chesnel}, \citenamefont {Sparks}, \citenamefont {Ward},
		\citenamefont {Yan}, \citenamefont {McGuire}, \citenamefont {Manley},
		\citenamefont {Staunton}, \citenamefont {Hermann},\ and\ \citenamefont
		{Frandsen}}]{baral;matter22}%
	\BibitemOpen
	\bibfield  {author} {\bibinfo {author} {\bibfnamefont {R.}~\bibnamefont
			{Baral}}, \bibinfo {author} {\bibfnamefont {J.}~\bibnamefont {Christensen}},
		\bibinfo {author} {\bibfnamefont {P.}~\bibnamefont {Hamilton}}, \bibinfo
		{author} {\bibfnamefont {F.}~\bibnamefont {Ye}}, \bibinfo {author}
		{\bibfnamefont {K.}~\bibnamefont {Chesnel}}, \bibinfo {author} {\bibfnamefont
			{T.~D.}\ \bibnamefont {Sparks}}, \bibinfo {author} {\bibfnamefont
			{R.}~\bibnamefont {Ward}}, \bibinfo {author} {\bibfnamefont {J.}~\bibnamefont
			{Yan}}, \bibinfo {author} {\bibfnamefont {M.~A.}\ \bibnamefont {McGuire}},
		\bibinfo {author} {\bibfnamefont {M.~E.}\ \bibnamefont {Manley}}, \bibinfo
		{author} {\bibfnamefont {J.~B.}\ \bibnamefont {Staunton}}, \bibinfo {author}
		{\bibfnamefont {R.~P.}\ \bibnamefont {Hermann}}, \ and\ \bibinfo {author}
		{\bibfnamefont {B.~A.}\ \bibnamefont {Frandsen}},\ }\bibfield  {title}
	{\enquote {\bibinfo {title} {Real-space visualization of short-range
				antiferromagnetic correlations in a magnetically enhanced thermoelectric},}\
	}\href {\doibase 10.1016/j.matt.2022.03.011} {\bibfield  {journal} {\bibinfo
			{journal} {Matter}\ }\textbf {\bibinfo {volume} {5}},\ \bibinfo {pages}
		{1853--1864} (\bibinfo {year} {2022})}\BibitemShut {NoStop}%
	\bibitem [{\citenamefont {Sch{\"a}rpf}\ and\ \citenamefont
		{Capellmann}(1993)}]{schar;pssa93}%
	\BibitemOpen
	\bibfield  {author} {\bibinfo {author} {\bibfnamefont {O.}~\bibnamefont
			{Sch{\"a}rpf}}\ and\ \bibinfo {author} {\bibfnamefont {H.}~\bibnamefont
			{Capellmann}},\ }\bibfield  {title} {\enquote {\bibinfo {title} {The
				xyz-difference method with polarized neutrons and the separation of coherent,
				spin incoherent, and magnetic scattering cross sections in a
				multidetector},}\ }\href@noop {} {\bibfield  {journal} {\bibinfo  {journal}
			{Phys. Status Solidi A}\ }\textbf {\bibinfo {volume} {135}},\ \bibinfo
		{pages} {359--379} (\bibinfo {year} {1993})}\BibitemShut {NoStop}%
	\bibitem [{\citenamefont {Billinge}\ and\ \citenamefont
		{Farrow}(2013)}]{billi;jpcm13}%
	\BibitemOpen
	\bibfield  {author} {\bibinfo {author} {\bibfnamefont {S.~J.~L.}\
			\bibnamefont {Billinge}}\ and\ \bibinfo {author} {\bibfnamefont {C.~L.}\
			\bibnamefont {Farrow}},\ }\bibfield  {title} {\enquote {\bibinfo {title}
			{Towards a robust \emph{ad-hoc} data correction approach that yields reliable
				atomic pair distribution functions from powder diffraction data},}\ }\href
	{\doibase 10.1088/0953-8984/25/45/454202} {\bibfield  {journal} {\bibinfo
			{journal} {J. Phys.: Condens. Mat.}\ }\textbf {\bibinfo {volume} {25}},\
		\bibinfo {pages} {454202} (\bibinfo {year} {2013})}\BibitemShut {NoStop}%
	\bibitem [{\citenamefont {Juh\'{a}s}\ \emph {et~al.}(2013)\citenamefont
		{Juh\'{a}s}, \citenamefont {Davis}, \citenamefont {Farrow},\ and\
		\citenamefont {Billinge}}]{juhas;jac13}%
	\BibitemOpen
	\bibfield  {author} {\bibinfo {author} {\bibfnamefont {P.}~\bibnamefont
			{Juh\'{a}s}}, \bibinfo {author} {\bibfnamefont {T.}~\bibnamefont {Davis}},
		\bibinfo {author} {\bibfnamefont {C.~L.}\ \bibnamefont {Farrow}}, \ and\
		\bibinfo {author} {\bibfnamefont {S.~J.~L.}\ \bibnamefont {Billinge}},\
	}\bibfield  {title} {\enquote {\bibinfo {title} {{{PDFgetX3}: A rapid and
					highly automatable program for processing powder diffraction data into total
					scattering pair distribution functions}},}\ }\href {\doibase
		10.1107/S0021889813005190} {\bibfield  {journal} {\bibinfo  {journal} {J.
				Appl. Crystallogr.}\ }\textbf {\bibinfo {volume} {46}},\ \bibinfo {pages}
		{560--566} (\bibinfo {year} {2013})}\BibitemShut {NoStop}%
	\bibitem [{\citenamefont {Frandsen}\ \emph {et~al.}(2022)\citenamefont
		{Frandsen}, \citenamefont {Parker}, \citenamefont {Christensen},
		\citenamefont {Stubben},\ and\ \citenamefont {Billinge}}]{frand;jac22}%
	\BibitemOpen
	\bibfield  {author} {\bibinfo {author} {\bibfnamefont {B.~A.}\ \bibnamefont
			{Frandsen}}, \bibinfo {author} {\bibfnamefont {H.~K.}\ \bibnamefont
			{Parker}}, \bibinfo {author} {\bibfnamefont {J.~A.}\ \bibnamefont
			{Christensen}}, \bibinfo {author} {\bibfnamefont {E.}~\bibnamefont
			{Stubben}}, \ and\ \bibinfo {author} {\bibfnamefont {S.~J.~L.}\ \bibnamefont
			{Billinge}},\ }\bibfield  {title} {\enquote {\bibinfo {title} {{diffpy.mpdf:
					open-source software for magnetic pair distribution function analysis}},}\
	}\href {\doibase 10.1107/S1600576722007257} {\bibfield  {journal} {\bibinfo
			{journal} {J. Appl. Crystallogr.}\ }\textbf {\bibinfo {volume} {55}},\
		\bibinfo {pages} {1377---1382} (\bibinfo {year} {2022})}\BibitemShut
	{NoStop}%
	\bibitem [{\citenamefont {Stewart}\ \emph {et~al.}(2009)\citenamefont
		{Stewart}, \citenamefont {Deen}, \citenamefont {Andersen}, \citenamefont
		{Schober}, \citenamefont {Barthelemy}, \citenamefont {Hillier}, \citenamefont
		{Murani}, \citenamefont {Hayes},\ and\ \citenamefont
		{Lindenau}}]{stewa;jac09}%
	\BibitemOpen
	\bibfield  {author} {\bibinfo {author} {\bibfnamefont {J.~R.}\ \bibnamefont
			{Stewart}}, \bibinfo {author} {\bibfnamefont {P.~P.}\ \bibnamefont {Deen}},
		\bibinfo {author} {\bibfnamefont {K.~H.}\ \bibnamefont {Andersen}}, \bibinfo
		{author} {\bibfnamefont {H.}~\bibnamefont {Schober}}, \bibinfo {author}
		{\bibfnamefont {J.-F.}\ \bibnamefont {Barthelemy}}, \bibinfo {author}
		{\bibfnamefont {J.~M.}\ \bibnamefont {Hillier}}, \bibinfo {author}
		{\bibfnamefont {A.~P.}\ \bibnamefont {Murani}}, \bibinfo {author}
		{\bibfnamefont {T.}~\bibnamefont {Hayes}}, \ and\ \bibinfo {author}
		{\bibfnamefont {B.}~\bibnamefont {Lindenau}},\ }\bibfield  {title} {\enquote
		{\bibinfo {title} {Disordered materials studied using neutron polarization
				analysis on the multi-detector spectrometer, d7},}\ }\href@noop {} {\bibfield
		{journal} {\bibinfo  {journal} {J. Appl. Crystallogr.}\ }\textbf {\bibinfo
			{volume} {42}},\ \bibinfo {pages} {69--84} (\bibinfo {year}
		{2009})}\BibitemShut {NoStop}%
	\bibitem [{\citenamefont {Ehlers}\ \emph {et~al.}(2013)\citenamefont {Ehlers},
		\citenamefont {Stewart}, \citenamefont {Wildes}, \citenamefont {Deen},\ and\
		\citenamefont {Andersen}}]{ehler;rsi13}%
	\BibitemOpen
	\bibfield  {author} {\bibinfo {author} {\bibfnamefont {G.}~\bibnamefont
			{Ehlers}}, \bibinfo {author} {\bibfnamefont {J.~R.}\ \bibnamefont {Stewart}},
		\bibinfo {author} {\bibfnamefont {A.~R.}\ \bibnamefont {Wildes}}, \bibinfo
		{author} {\bibfnamefont {P.~P.}\ \bibnamefont {Deen}}, \ and\ \bibinfo
		{author} {\bibfnamefont {K.~H.}\ \bibnamefont {Andersen}},\ }\bibfield
	{title} {\enquote {\bibinfo {title} {Generalization of the classical
				xyz-polarization analysis technique to out-of-plane and inelastic
				scattering},}\ }\href@noop {} {\bibfield  {journal} {\bibinfo  {journal}
			{Rev. Sci. Instrum.}\ }\textbf {\bibinfo {volume} {84}},\ \bibinfo {pages}
		{093901} (\bibinfo {year} {2013})}\BibitemShut {NoStop}%
	\bibitem [{\citenamefont {Zentrum}(2015)}]{hmlz;jlsrf15}%
	\BibitemOpen
	\bibfield  {author} {\bibinfo {author} {\bibfnamefont {H.~M.-L.}\
			\bibnamefont {Zentrum}},\ }\bibfield  {title} {\enquote {\bibinfo {title}
			{{DNS: Diffuse scattering neutron time-of-flight spectrometer}},}\ }\href
	{\doibase 10.17815/jlsrf-1-33} {\bibfield  {journal} {\bibinfo  {journal}
			{JLSRF}\ }\textbf {\bibinfo {volume} {1}},\ \bibinfo {pages} {A27} (\bibinfo
		{year} {2015})}\BibitemShut {NoStop}%
	\bibitem [{\citenamefont {Zaliznyak}\ \emph {et~al.}(2017)\citenamefont
		{Zaliznyak}, \citenamefont {Savici}, \citenamefont {Garlea}, \citenamefont
		{Winn}, \citenamefont {Filges}, \citenamefont {Schneeloch}, \citenamefont
		{Tranquada}, \citenamefont {Gu}, \citenamefont {Wang},\ and\ \citenamefont
		{Petrovic}}]{zaliz;jpconfs17}%
	\BibitemOpen
	\bibfield  {author} {\bibinfo {author} {\bibfnamefont {I.~A.}\ \bibnamefont
			{Zaliznyak}}, \bibinfo {author} {\bibfnamefont {A.~T.}\ \bibnamefont
			{Savici}}, \bibinfo {author} {\bibfnamefont {V.~O.}\ \bibnamefont {Garlea}},
		\bibinfo {author} {\bibfnamefont {B.}~\bibnamefont {Winn}}, \bibinfo {author}
		{\bibfnamefont {U.}~\bibnamefont {Filges}}, \bibinfo {author} {\bibfnamefont
			{J.}~\bibnamefont {Schneeloch}}, \bibinfo {author} {\bibfnamefont {J.~M.}\
			\bibnamefont {Tranquada}}, \bibinfo {author} {\bibfnamefont {G.}~\bibnamefont
			{Gu}}, \bibinfo {author} {\bibfnamefont {A.}~\bibnamefont {Wang}}, \ and\
		\bibinfo {author} {\bibfnamefont {C.}~\bibnamefont {Petrovic}},\ }\bibfield
	{title} {\enquote {\bibinfo {title} {Polarized neutron scattering on
				{HYSPEC}: the {HYbrid} {SPECtrometer} at {SNS}},}\ }\href {\doibase
		10.1088/1742-6596/862/1/012030} {\bibfield  {journal} {\bibinfo  {journal}
			{J. Phys.: Conf. Ser.}\ }\textbf {\bibinfo {volume} {862}},\ \bibinfo {pages}
		{012030} (\bibinfo {year} {2017})}\BibitemShut {NoStop}%
	\bibitem [{\citenamefont {Kunitomi}, \citenamefont {Hamaguchi},\ and\
		\citenamefont {Anzai}(1964)}]{kunit;jdp64}%
	\BibitemOpen
	\bibfield  {author} {\bibinfo {author} {\bibfnamefont {N.}~\bibnamefont
			{Kunitomi}}, \bibinfo {author} {\bibfnamefont {Y.}~\bibnamefont {Hamaguchi}},
		\ and\ \bibinfo {author} {\bibfnamefont {S.}~\bibnamefont {Anzai}},\
	}\bibfield  {title} {\enquote {\bibinfo {title} {{Neutron diffraction study
					on manganese telluride}},}\ }\href {\doibase 10.1051/jphys:01964002505056800}
	{\bibfield  {journal} {\bibinfo  {journal} {J. Phys.-Paris}\ }\textbf
		{\bibinfo {volume} {25}},\ \bibinfo {pages} {568--574} (\bibinfo {year}
		{1964})}\BibitemShut {NoStop}%
	\bibitem [{\citenamefont {D'Sa}\ \emph {et~al.}(2005)\citenamefont {D'Sa},
		\citenamefont {Bhobe}, \citenamefont {Priolkar}, \citenamefont {Das},
		\citenamefont {Paranjpe}, \citenamefont {Prabhu},\ and\ \citenamefont
		{Sarode}}]{dsa;jmmm05}%
	\BibitemOpen
	\bibfield  {author} {\bibinfo {author} {\bibfnamefont {J.~E.}\ \bibnamefont
			{D'Sa}}, \bibinfo {author} {\bibfnamefont {P.}~\bibnamefont {Bhobe}},
		\bibinfo {author} {\bibfnamefont {K.}~\bibnamefont {Priolkar}}, \bibinfo
		{author} {\bibfnamefont {A.}~\bibnamefont {Das}}, \bibinfo {author}
		{\bibfnamefont {S.}~\bibnamefont {Paranjpe}}, \bibinfo {author}
		{\bibfnamefont {R.}~\bibnamefont {Prabhu}}, \ and\ \bibinfo {author}
		{\bibfnamefont {P.}~\bibnamefont {Sarode}},\ }\bibfield  {title} {\enquote
		{\bibinfo {title} {Low-temperature neutron diffraction study of mnte},}\
	}\href@noop {} {\bibfield  {journal} {\bibinfo  {journal} {J. Magn. Magn.
				Mater.}\ }\textbf {\bibinfo {volume} {285}},\ \bibinfo {pages} {267--271}
		(\bibinfo {year} {2005})}\BibitemShut {NoStop}%
	\bibitem [{\citenamefont {Szuszkiewicz}\ \emph {et~al.}(2006)\citenamefont
		{Szuszkiewicz}, \citenamefont {Dynowska}, \citenamefont {Witkowska},\ and\
		\citenamefont {Hennion}}]{szusz;prb06}%
	\BibitemOpen
	\bibfield  {author} {\bibinfo {author} {\bibfnamefont {W.}~\bibnamefont
			{Szuszkiewicz}}, \bibinfo {author} {\bibfnamefont {E.}~\bibnamefont
			{Dynowska}}, \bibinfo {author} {\bibfnamefont {B.}~\bibnamefont {Witkowska}},
		\ and\ \bibinfo {author} {\bibfnamefont {B.}~\bibnamefont {Hennion}},\
	}\bibfield  {title} {\enquote {\bibinfo {title} {{Spin-wave measurements on
					hexagonal MnTe NiAs-type structure by inelastic neutron scattering}},}\
	}\href {\doibase 10.1103/PhysRevB.73.104403} {\bibfield  {journal} {\bibinfo
			{journal} {Phys. Rev. B}\ }\textbf {\bibinfo {volume} {73}},\ \bibinfo
		{pages} {104403} (\bibinfo {year} {2006})}\BibitemShut {NoStop}%
	\bibitem [{\citenamefont {Zheng}\ \emph {et~al.}(2019)\citenamefont {Zheng},
		\citenamefont {Lu}, \citenamefont {Polash}, \citenamefont
		{Rasoulianboroujeni}, \citenamefont {Liu}, \citenamefont {Manley},
		\citenamefont {Deng}, \citenamefont {Sun}, \citenamefont {Chen},
		\citenamefont {Hermann}, \citenamefont {Vashaee}, \citenamefont {Heremans},\
		and\ \citenamefont {Zhao}}]{zheng;sadv19}%
	\BibitemOpen
	\bibfield  {author} {\bibinfo {author} {\bibfnamefont {Y.}~\bibnamefont
			{Zheng}}, \bibinfo {author} {\bibfnamefont {T.}~\bibnamefont {Lu}}, \bibinfo
		{author} {\bibfnamefont {M.~M.~H.}\ \bibnamefont {Polash}}, \bibinfo {author}
		{\bibfnamefont {M.}~\bibnamefont {Rasoulianboroujeni}}, \bibinfo {author}
		{\bibfnamefont {N.}~\bibnamefont {Liu}}, \bibinfo {author} {\bibfnamefont
			{M.~E.}\ \bibnamefont {Manley}}, \bibinfo {author} {\bibfnamefont
			{Y.}~\bibnamefont {Deng}}, \bibinfo {author} {\bibfnamefont {P.~J.}\
			\bibnamefont {Sun}}, \bibinfo {author} {\bibfnamefont {X.~L.}\ \bibnamefont
			{Chen}}, \bibinfo {author} {\bibfnamefont {R.~P.}\ \bibnamefont {Hermann}},
		\bibinfo {author} {\bibfnamefont {D.}~\bibnamefont {Vashaee}}, \bibinfo
		{author} {\bibfnamefont {J.~P.}\ \bibnamefont {Heremans}}, \ and\ \bibinfo
		{author} {\bibfnamefont {H.}~\bibnamefont {Zhao}},\ }\bibfield  {title}
	{\enquote {\bibinfo {title} {{Paramagnon drag in high thermoelectric figure
					of merit Li-doped MnTe}},}\ }\href {\doibase 10.1126/sciadv.aat9461}
	{\bibfield  {journal} {\bibinfo  {journal} {Sci. Adv.}\ }\textbf {\bibinfo
			{volume} {5}},\ \bibinfo {pages} {eaat9461} (\bibinfo {year}
		{2019})}\BibitemShut {NoStop}%
	\bibitem [{\citenamefont {Roth}\ \emph {et~al.}(2018)\citenamefont {Roth},
		\citenamefont {May}, \citenamefont {Ye}, \citenamefont {Chakoumakos},\ and\
		\citenamefont {Iversen}}]{roth;iucrj18}%
	\BibitemOpen
	\bibfield  {author} {\bibinfo {author} {\bibfnamefont {N.}~\bibnamefont
			{Roth}}, \bibinfo {author} {\bibfnamefont {A.~F.}\ \bibnamefont {May}},
		\bibinfo {author} {\bibfnamefont {F.}~\bibnamefont {Ye}}, \bibinfo {author}
		{\bibfnamefont {B.~C.}\ \bibnamefont {Chakoumakos}}, \ and\ \bibinfo {author}
		{\bibfnamefont {B.~B.}\ \bibnamefont {Iversen}},\ }\bibfield  {title}
	{\enquote {\bibinfo {title} {Model-free reconstruction of magnetic
				correlations in frustrated magnets},}\ }\href@noop {} {\bibfield  {journal}
		{\bibinfo  {journal} {IUCrJ}\ }\textbf {\bibinfo {volume} {5}},\ \bibinfo
		{pages} {410--416} (\bibinfo {year} {2018})}\BibitemShut {NoStop}%
	\bibitem [{\citenamefont {Savici}\ \emph {et~al.}(2017)\citenamefont {Savici},
		\citenamefont {Zaliznyak}, \citenamefont {Garlea},\ and\ \citenamefont
		{Winn}}]{savic;jpconfs17}%
	\BibitemOpen
	\bibfield  {author} {\bibinfo {author} {\bibfnamefont {A.~T.}\ \bibnamefont
			{Savici}}, \bibinfo {author} {\bibfnamefont {I.~A.}\ \bibnamefont
			{Zaliznyak}}, \bibinfo {author} {\bibfnamefont {V.~O.}\ \bibnamefont
			{Garlea}}, \ and\ \bibinfo {author} {\bibfnamefont {B.}~\bibnamefont
			{Winn}},\ }\bibfield  {title} {\enquote {\bibinfo {title} {Data processing
				workflow for time of flight polarized neutrons inelastic measurements},}\
	}\href {\doibase 10.1088/1742-6596/862/1/012023} {\bibfield  {journal}
		{\bibinfo  {journal} {J. Phys.: Conf. Ser.}\ }\textbf {\bibinfo {volume}
			{862}},\ \bibinfo {pages} {012023} (\bibinfo {year} {2017})}\BibitemShut
	{NoStop}%
	\bibitem [{\citenamefont {Wilson}(1995)}]{wilso;b;itc95}%
	\BibitemOpen
	\bibinfo {editor} {\bibfnamefont {A.}~\bibnamefont {Wilson}},\ ed.,\
	\href@noop {} {\emph {\bibinfo {title} {International Tables for
				Crystallography, Volume C: Mathematical, Physical, and Chemical Tables}}}\
	(\bibinfo  {publisher} {Kluwer},\ \bibinfo {address} {Dordrecht, The
		Netherlands},\ \bibinfo {year} {1995})\BibitemShut {NoStop}%
	\bibitem [{\citenamefont {Farrow}\ \emph {et~al.}(2007)\citenamefont {Farrow},
		\citenamefont {Juh\'as}, \citenamefont {Liu}, \citenamefont {Bryndin},
		\citenamefont {{Bo\v zin}}, \citenamefont {Bloch}, \citenamefont {Proffen},\
		and\ \citenamefont {Billinge}}]{farro;jpcm07}%
	\BibitemOpen
	\bibfield  {author} {\bibinfo {author} {\bibfnamefont {C.~L.}\ \bibnamefont
			{Farrow}}, \bibinfo {author} {\bibfnamefont {P.}~\bibnamefont {Juh\'as}},
		\bibinfo {author} {\bibfnamefont {J.}~\bibnamefont {Liu}}, \bibinfo {author}
		{\bibfnamefont {D.}~\bibnamefont {Bryndin}}, \bibinfo {author} {\bibfnamefont
			{E.~S.}\ \bibnamefont {{Bo\v zin}}}, \bibinfo {author} {\bibfnamefont
			{J.}~\bibnamefont {Bloch}}, \bibinfo {author} {\bibfnamefont
			{T.}~\bibnamefont {Proffen}}, \ and\ \bibinfo {author} {\bibfnamefont
			{S.~J.~L.}\ \bibnamefont {Billinge}},\ }\bibfield  {title} {\enquote
		{\bibinfo {title} {{PDFfit2} and {PDFgui}: Computer programs for studying
				nanostructure in crystals},}\ }\href {\doibase
		10.1088/0953-8984/19/33/335219} {\bibfield  {journal} {\bibinfo  {journal}
			{J. Phys.: Condens. Mat.}\ }\textbf {\bibinfo {volume} {19}},\ \bibinfo
		{pages} {335219} (\bibinfo {year} {2007})}\BibitemShut {NoStop}%
	\bibitem [{\citenamefont {Kodama}\ \emph {et~al.}(2021)\citenamefont {Kodama},
		\citenamefont {Honda}, \citenamefont {Ikeda}, \citenamefont {Shamoto},\ and\
		\citenamefont {Otomo}}]{kodam;jpscp21}%
	\BibitemOpen
	\bibfield  {author} {\bibinfo {author} {\bibfnamefont {K.}~\bibnamefont
			{Kodama}}, \bibinfo {author} {\bibfnamefont {T.}~\bibnamefont {Honda}},
		\bibinfo {author} {\bibfnamefont {K.}~\bibnamefont {Ikeda}}, \bibinfo
		{author} {\bibfnamefont {S.-i.}\ \bibnamefont {Shamoto}}, \ and\ \bibinfo
		{author} {\bibfnamefont {T.}~\bibnamefont {Otomo}},\ }\bibfield  {title}
	{\enquote {\bibinfo {title} {{Magnetic pair distribution function of
					spin-glass system Mn$_{0.5}$Fe$_{0.5}$TiO$_{3}$}},}\ }\href {\doibase
		10.7566/JPSCP.33.011059} {\bibfield  {journal} {\bibinfo  {journal} {JPS
				Conf. Proc.}\ }\textbf {\bibinfo {volume} {33}},\ \bibinfo {pages} {011059}
		(\bibinfo {year} {2021})}\BibitemShut {NoStop}%
	\bibitem [{\citenamefont {Soper}\ and\ \citenamefont
		{Barney}(2012)}]{soper;jac12}%
	\BibitemOpen
	\bibfield  {author} {\bibinfo {author} {\bibfnamefont {A.~K.}\ \bibnamefont
			{Soper}}\ and\ \bibinfo {author} {\bibfnamefont {E.~R.}\ \bibnamefont
			{Barney}},\ }\bibfield  {title} {\enquote {\bibinfo {title} {{On the use of
					modification functions when Fourier transforming total scattering data}},}\
	}\href {\doibase 10.1107/S002188981203960X} {\bibfield  {journal} {\bibinfo
			{journal} {J. Appl. Crystallogr.}\ }\textbf {\bibinfo {volume} {45}},\
		\bibinfo {pages} {1314--1317} (\bibinfo {year} {2012})}\BibitemShut {NoStop}%
	\bibitem [{\citenamefont {Lorch}(1969)}]{lorch;jpcss69}%
	\BibitemOpen
	\bibfield  {author} {\bibinfo {author} {\bibfnamefont {E.}~\bibnamefont
			{Lorch}},\ }\bibfield  {title} {\enquote {\bibinfo {title} {{Neutron
					diffraction by germania, silica and radiation-damaged silica glasses}},}\
	}\href {\doibase 10.1088/0022-3719/2/2/305} {\bibfield  {journal} {\bibinfo
			{journal} {J. Phys. C: Solid State Phys.}\ }\textbf {\bibinfo {volume} {2}},\
		\bibinfo {pages} {229} (\bibinfo {year} {1969})}\BibitemShut {NoStop}%
	\bibitem [{\citenamefont {Garlea}\ \emph {et~al.}(2022)\citenamefont {Garlea},
		\citenamefont {Calder}, \citenamefont {Huegle}, \citenamefont {Lin},
		\citenamefont {Islam}, \citenamefont {Stoica}, \citenamefont {Graves},
		\citenamefont {Frandsen},\ and\ \citenamefont {Wilson}}]{garle;rsi22}%
	\BibitemOpen
	\bibfield  {author} {\bibinfo {author} {\bibfnamefont {V.~O.}\ \bibnamefont
			{Garlea}}, \bibinfo {author} {\bibfnamefont {S.}~\bibnamefont {Calder}},
		\bibinfo {author} {\bibfnamefont {T.}~\bibnamefont {Huegle}}, \bibinfo
		{author} {\bibfnamefont {J.~Y.~Y.}\ \bibnamefont {Lin}}, \bibinfo {author}
		{\bibfnamefont {F.}~\bibnamefont {Islam}}, \bibinfo {author} {\bibfnamefont
			{A.}~\bibnamefont {Stoica}}, \bibinfo {author} {\bibfnamefont {V.~B.}\
			\bibnamefont {Graves}}, \bibinfo {author} {\bibfnamefont {B.}~\bibnamefont
			{Frandsen}}, \ and\ \bibinfo {author} {\bibfnamefont {S.~D.}\ \bibnamefont
			{Wilson}},\ }\bibfield  {title} {\enquote {\bibinfo {title} {{VERDI:
					VERsatile DIffractometer with wide-angle polarization analysis for magnetic
					structure studies in powders and single crystals}},}\ }\href {\doibase
		10.1063/5.0090919} {\bibfield  {journal} {\bibinfo  {journal} {Rev. Sci.
				Instrum.}\ }\textbf {\bibinfo {volume} {93}},\ \bibinfo {pages} {065103}
		(\bibinfo {year} {2022})}\BibitemShut {NoStop}%
\end{thebibliography}
\end{document}